\documentclass[fleqn,usenatbib]{mnras}

\usepackage[T1]{fontenc}

\DeclareRobustCommand{\VAN}[3]{#2}
\let\VANthebibliography\thebibliography
\def\thebibliography{\DeclareRobustCommand{\VAN}[3]{##3}\VANthebibliography}

\usepackage{graphicx}	
\usepackage{amsmath}	
\usepackage{amssymb}	
\usepackage{bm}
\usepackage{comment}
\graphicspath{{figures/}}
\usepackage{newtxtext,newtxmath}

\newcommand{\logp}{\log_{10}(P_\mathrm{rot}/\mathrm{day})}	

\newcommand{\prot}{P_\mathrm{rot}}

\title[Rotation period distribution from $v\sin i$]{
Inferring the Rotation Period Distribution of Stars
from their Projected Rotation Velocities and Radii: Application to late-F/early-G \textit{Kepler} Stars}

\author[Masuda et al.]{
Kento Masuda$^{1}$\thanks{E-mail: kmasuda@ess.sci.osaka-u.ac.jp},
Erik A.\ Petigura$^{2}$, and
Oliver J. Hall$^{3}$
\\
$^{1}$Department of Earth and Space Science, Osaka University, Osaka 560-0043, Japan
\\
$^{2}$Department of Physics \& Astronomy, University of California Los Angeles, Los Angeles, CA 90095, USA
\\
$^{3}$European Space Agency (ESA), European Space Research and Technology Centre (ESTEC), Keplerlaan 1, 2201 AZ Noordwijk, The Netherlands
}

\date{Accepted XXX. Received YYY; in original form ZZZ}

\pubyear{2021}

\begin{document}
\label{firstpage}
\pagerange{\pageref{firstpage}--\pageref{lastpage}}
\maketitle

\begin{abstract}

While stellar rotation periods $P_\mathrm{rot}$ may be measured from broadband photometry, the photometric modulation becomes harder to detect for slower rotators, which could bias measurements of the long-period tail of the $P_\mathrm{rot}$ distribution. Alternatively, the $P_\mathrm{rot}$ distribution of stars can be inferred from their projected rotation velocities $v\sin i$ and radii $R$, without being biased against photometrically quiet stars. We solve this inference problem using a hierarchical Bayesian framework, which (i) is applicable to heteroscedastic measurements of $v\sin i$ and $R$ with non-Gaussian uncertainties and (ii) does not require a simple parametric form for the true $P_\mathrm{rot}$ distribution. We test the method on simulated data sets and show that the true $P_\mathrm{rot}$ distribution can be recovered from $\gtrsim 100$ sets of $v\sin i$ and $R$ measured with precisions of $1\,\mathrm{km/s}$ and $4\%$, respectively, unless the true distribution includes sharp discontinuities. We apply the method to a sample of 144 late-F/early-G dwarfs in the \textit{Kepler} field with $v\sin i$ measured from Keck/HIRES spectra, and find that the typical rotation periods of these stars are similar to the photometric periods measured from \textit{Kepler} light curves: we do not find a large population of slow rotators that are missed in the photometric sample, although we find evidence that the photometric sample is biased for young, rapidly-rotating stars. Our results also agree with asteroseismic measurements of $P_\mathrm{rot}$ for \textit{Kepler} stars with similar ages and effective temperatures, and show that $\approx 1.1\,M_\odot$ stars beyond the middle of their main-sequence lifetimes rotate faster than predicted by standard magnetic braking laws.

\end{abstract}




\section{Introduction}

The study of stellar rotation has been revolutionized by high-precision, continuous, and long-term photometry made available by the \textit{Kepler} mission \citep{2010Sci...327..977B, 2010ApJ...713L..79K}. Quasi-periodic brightness variations due to active regions on the surface enabled rotation period measurements for tens of thousands of FGKM stars, including those with longer-period and/or lower-amplitude variabilities than can be studied with the ground-based photometry \citep[e.g.][]{2013A&A...557L..10N, 2013A&A...560A...4R, 2014ApJS..211...24M, 2014A&A...572A..34G,2019ApJS..244...21S,2020A&A...635A..43R}. The data from the ongoing space missions such as \textit{TESS} \citep{2014SPIE.9143E..20R} and \textit{Gaia} \citep{2016A&A...595A...1G} will further expand the capability of the photometry-based measurements \citep[e.g.][]{2018A&A...616A..16L,2021arXiv210414566C}.

That said, the photometric method is not without limitations. For Sun-like stars, the fraction of stars with robustly detected rotational modulation is less than about a third even in the \textit{Kepler} data \citep{2014ApJS..211...24M,2021ApJS..255...17S}; no information is available for the majority of the sample.
The selection function for the period detection is difficult to quantify, and the detectability of rotational modulation is also likely correlated with various properties including the stellar age, rotation period, spin inclination, and surface distribution of active regions. These are often the exact features that one wishes to study, and so this unknown selection function complicates the interpretation of ensemble studies \citep[see, e.g.,][]{2015ApJ...801....3M, 2019A&A...621A..21R, 2021ApJ...908L..21R}.
Even when a periodicity is firmly detected, it is not a trivial task to distinguish rotational modulation from other astrophysical variabilities, and the relation between the detected periodicity and the true one (e.g., the detected period may correspond to harmonics of the true period) depends on the method to analyze the frequency components of the light curve as well as
the unknown morphology of surface active regions \citep[e.g.][]{2013MNRAS.432.1203M}. Indeed, individual photometric measurements are not always in good agreement with presumably more reliable measurements using rotational splitting of asteroseismic oscillation modes \citep{2018MNRAS.479..391K, 2019AJ....157..172S}.

These difficulties motivate us to study an alternative method using the projected rotational velocity $v\sin i$ measured from the absorption line broadening \citep[e.g.][]{1966ApJ...144..695W, 1967ApJ...150..551K, 1972ApJ...171..565S}. Ignoring the differential rotation, 
\begin{equation}
    v\sin i = {2\pi R \over \prot}\,\sin i
\end{equation}
can be ``solved for'' the rotation period $\prot$ --- if the radius $R$ is known from other measurements and spin orientations are known to be isotropic (i.e., $\cos i$ has a uniform distribution, where $i$ is the inclination of the spin axis measured from our line of sight).
A key advantage of the $v\sin i$ method is that it is not biased against less active stars that may not exhibit detectable photometric brightness variations. 
Although the method requires high-resolution spectra and precise measurements of stellar radii, spectroscopic surveys delivering high-resolution spectra for millions of stars, combined with parallax information from \textit{Gaia}, will overcome those limitations of the method and significantly expand its applicability.

A challenge to this inference is that the $\prot$ of individual stars cannot be precisely measured due to unknown $\sin i$ and that one needs to work on a distribution of $\prot$ for a sample of many stars. 
This ``rectification'' of the $\sin i$ factor is itself a solved problem: assuming isotropic spin orientations, the relation between the distributions of $q \equiv v \sin i/2\pi R$ and that of $q/\sin i=1/\prot$ can be reduced to Abel's integral equation that can be solved analytically \citep{1950ApJ...111..142C}. This involves differentiation of the probability density function (PDF) for $q$, and so \citet{1950ApJ...111..142C} remarked that one should focus on moments, or to assume simple (smooth) functional forms for the PDF $f(q)$. An alternative approach is to adopt an iterative algorithm widely known as Lucy-Richardson deconvolution \citep{1972JOSA...62...55R, 1974AJ.....79..745L}. This can be used to construct a smooth distribution of $q/\sin i$ from $f(q)$, but the results can be sensitive to how one constructs $f(q)$ from, e.g., a noisy histogram of $q$. In short, these methods {\it require} that the $q$ distribution is well estimated from the data,
and do not provide a means to {\it infer} $f(q)$. One needs to obtain a sensible $f(q)$ for successful application of these traditional methods. 

Inferring $f(q)$ is not simple when the number of measurements is small and/or $v\sin i$ is comparable to its uncertainty. In the former case, the measured $q$ may only sparsely cover the support of $f(q)$, and in the latter case, the measurements are often heteroscedastic and have 
non-Gaussian uncertainties (e.g., it is often the case that only upper limits can be reliably estimated for Sun-like stars). When one is interested in rotation periods (rather than the equatorial rotation velocity $v$ as in the traditional applications), the uncertainty of $R$ also needs to be taken into account. 
Here we develop a framework that simultaneously accounts for measurement and geometrical uncertainties to infer the stellar rotation period distribution with minimal assumptions on its functional form.\footnote{\citet{2007A&A...463..671R} employed the kernel density estimation to infer $f(q)$ \citep[see also][]{2001A&A...379..992J}. They then proceeded to perform multiple deconvolutions using Lucy's method, first to correct for measurement errors (which in their case could be assumed to be homoscedastic) and then for the $\sin i$ effect. It may be possible to generalize such a procedure for heteroscedatsic and non-Gaussian measurements but the resulting procedure will become even more complicated, especially if one also considers the division by $R$ to discuss rotation periods.
Here we seek for an alternative approach that makes it more straightforward to take into account all these effects.
}

In Section \ref{sec:method}, we present a mathematical framework. In Section \ref{sec:tests}, we apply the method to simulated data sets and illustrate 
the situations where our method succeeds and where it fails.
In Section \ref{sec:kepler}, we apply the method to actual measurements of $v\sin i$ obtained from Keck/HIRES spectra for late-F/early-G stars in the \textit{Kepler} field to infer the rotation period distribution that is not conditioned on the detectable photometric modulation.
We compare the result with rotation periods measured using photometric variations and asteroseismology, and discuss its implications for the spin-down law of near Solar-mass stars. Section \ref{sec:summary} summarizes and concludes the paper.

\section{The Method}\label{sec:method}

Ignoring differential rotation, individual measurements of $v\sin i$ and $R$ constrain
\begin{equation}
    {v\sin i \over {2\pi R}} = {1 \over \prot}\sin i.
\end{equation}
The inclination $i$ is not known for individual stars, but its PDF is known for a sample of stars with isotropically oriented spins: $p(i)d i \propto \sin i d i$. We wish to infer $p(\prot)$ in this sample based on the measurements (i.e., likelihood functions) of $v\sin i$ and $R$ for individual stars and the assumed $p(i)$. The formulation largely follows the approach proposed by \citet{2010ApJ...725.2166H} for inferring the exoplanet eccentricity distribution.

\subsection{Assumptions} \label{ssec:method_assumptions}

For each star labelled by $j$, we have the data $D_{u,j}$ and $D_{R,j}$ that provide likelihood functions for $u\equiv v \sin i$ and $R$. The forms of the likelihood functions can be arbitrary, and we assume that all the $D_{u,j}$ and $D_{R,j}$ are independent. We assume that spins are isotropically oriented and are uncorrelated with any other properties of the stars including their radii: the prior PDF for $\cos i$ of each star is uniform and is independent from $R$ (whose prior PDF is also assumed to be uniform and separable). Rotation periods of the sample stars are assumed to be drawn from the common underlying distribution (hyperprior), $p(\prot|\alpha)$, where $\alpha$ is a set of parameters (hyperparameters) that control the shape of the distribution. 
Our problem then reduces to inferring $\alpha$ from a set of observational data $D=\{D_{u,j}, D_{R,j}\}$. 

Here we also assume that the hyperprior $p(\prot|\alpha)$ is independent from any other stellar parameters including $R$ and $i$. This may not be true in general: $\prot$ distribution depends on stellar types, and hence $R$. In this paper, we do not discuss such a generalization, but do examine dependence on the effective temperatures of the sample stars in Section \ref{sec:kepler}. 

\subsection{Hierarchical Framework}\label{ssec:method_framework}

Given the set of assumptions in Section \ref{ssec:method_assumptions}, 
the likelihood for the data sets of $N$ stars is
\begin{align}
\notag
    p(D|\alpha) 
    &= \prod_{j=1}^N p(D_{u,j}, D_{R,j}|\alpha)\\
\label{eq:likelihood}
    &= \prod_{j=1}^N \int_{x_{\rm min}}^{x_{\rm max}} p(D_{u,j}, D_{R,j}|x)\,p(x|\alpha)\,\mathrm{d}x,
\end{align}
where $x=\logp$, and $(x_{\rm min}, x_{\rm max})$ are the prior bounds specified below (see Appendix \ref{sec:derivation} for more detailed derivation of the equations in this section). Here the marginal likelihood for $x_j$ is
\begin{align}
\notag
    &p(D_{u,j}, D_{R,j}|x_j) \\
\notag
    &= \int p(D_{u,j}, D_{R,j}|P_{\mathrm{rot},j}, R_j, \cos i_j)\,p(R_j)\,p(\cos i_j)\,\mathrm{d}R_j\,\mathrm{d}\cos i_j \\
\label{eq:margL}
    &= \int p\left(D_{u,j}|{2\pi R_j \over P_{\mathrm{rot},j}}\sqrt{1-\cos^2 i_j}\right)\,p(D_{R,j}|R_j)\,p(R_j)\,p(\cos i_j)\,\mathrm{d}R_j\,\mathrm{d}\cos i_j, 
\end{align}
where integration is performed over $(0,\infty)$ for $R_j$ and $(0,1)$ for $\cos i_j$, respectively.
Given the measurements of $R$ and $v\sin i$, this integral can be computed for each star once. When we infer $\alpha$, Equation \ref{eq:likelihood} is computed from $p(x|\alpha)$ and the pre-computed values of $p(D_{u,j}, D_{R,j}|x)$. 

We model the $p(x|\alpha)$ as a step function:
\begin{equation}
    p(x|\alpha) = \sum_{m=1}^{M} \exp(\alpha_m)\,\Pi \left(x; x_\mathrm{min}+(m-1)\Delta x, x_\mathrm{min}+m\Delta x\right), 
\end{equation}
where 
\begin{equation}
    \Pi(x; a,b) = \begin{cases}
    0, \quad &x<a \quad \mathrm{or}\quad x>b\\
    {1\over{b-a}}, \quad &a\leq x \leq b
    \end{cases}
\end{equation}
and $\Delta x = {{x_\mathrm{max}-x_\mathrm{min}} \over M}$. In other words, we model $p(x|\alpha)$ as a histogram with $M$ bins whose logarithmic bin heights $\{\alpha_m\}_{m=1}^M$ are the parameters to be inferred.

To obtain $p(\alpha|D)$, we also need to assign a prior PDF for $\alpha$. To ensure the smoothness that is naturally expected, we employ a Gaussian process \citep{2006gpml.book.....R, 2014ApJ...795...64F}. We also require the sum $\sum_m \exp(\alpha_m)\,\Delta x$ to be unity so that $p(x|\alpha)$ is a normalized PDF for $x$. The latter was ensured by first sampling $\beta$ from an $M$-dimensional normal distribution $\mathcal{N}(\beta; 0, \sigma_\beta^2)$ and by converting it to
\begin{equation}
    \alpha = \beta - \ln\left[\sum_{m=1}^M \exp(\beta_m)\Delta x\right].
\end{equation} 
This operation implies a certain form of the PDF for $\alpha$ which we denote by $\pi (\alpha)$. Then the full prior on $\alpha$ is:
\begin{equation}
    p(\alpha|\ln a, \ln s) \propto \pi(\alpha)\,\mathcal{N}(\alpha; \overline{\alpha}, K(a, s)),
\end{equation}
where $\overline{\alpha}$ is fixed to be the mean value that gives flat and normalized $p(x|\alpha)$, and 
\begin{equation}
    K_{ij}(a,s) = a^2 \left(1 + {\sqrt{3}|i - j|\Delta x\over {s}}\right)\,\exp\left(-{\sqrt{3}|i - j|\Delta x \over {s}} \right)
\end{equation}
is a Mat\'{e}rn-3/2 covariance function \citep{2006gpml.book.....R}. Here the parameters $a$ and $s$ represent the strength and length scale of the correlation between different bins, respectively.

In this paper, we choose $x_\mathrm{min}=0$ and $x_\mathrm{max}=2$ ($1$--$100\,\mathrm{days}$), $M=100$, $\sigma_\beta=10$, the $\ln a$ prior uniform between $(-5,5)$, and the $\ln s$ prior uniform between $(-2, 2)$.
This choice is not unique nor physically motivated, 
but was found to 
recover the period distribution successfully when applied to simulated data sets (Section \ref{sec:tests}) that have similar properties to the actual data we discuss in Section \ref{sec:kepler}.
When applied to data sets with different sizes and qualities, the choice of the prior (including the form of the hyperprior) may need to be revisited using similar simulations.

Using this hyperprior, we sample from
\begin{equation}
    p(\alpha, \ln a, \ln s |D) \propto p(D|\alpha)\,p(\alpha|\ln a, \ln s)\,p(\ln a)\,p(\ln s)
\end{equation}
and marginalize over $\ln a$ and $\ln s$ to obtain $p(\alpha|D)$. Then the
period distribution conditioned on all data is computed as
\begin{equation}
    \label{eq:meanposterior}
    p(x|D) = \int p(x|\alpha)\,p(\alpha|D)\,\mathrm{d}\alpha.
\end{equation}

The code was implemented and tested using {\tt JAX} 0.2.12 \citep{jax2018github} and {\tt NumPyro} 0.4.0 \citep{bingham2018pyro, phan2019composable}. The posterior samples for the parameters ${\alpha}$, $\ln a$, and $\ln s$ were obtained using Hamiltonian Monte Carlo \citep{DUANE1987216, 2017arXiv170102434B} with No-U-Turn sampler \citep{2011arXiv1111.4246H}. 
Computation of the Gaussian-process likelihood was performed using {\tt celerite2} \citep{celerite1, celerite2}.
We sampled until the resulting chains had the split $\hat{R}<1.02$ \citep{BB13945229} for all the parameters. 

\begin{figure*}
	\includegraphics[width=1.26\columnwidth]{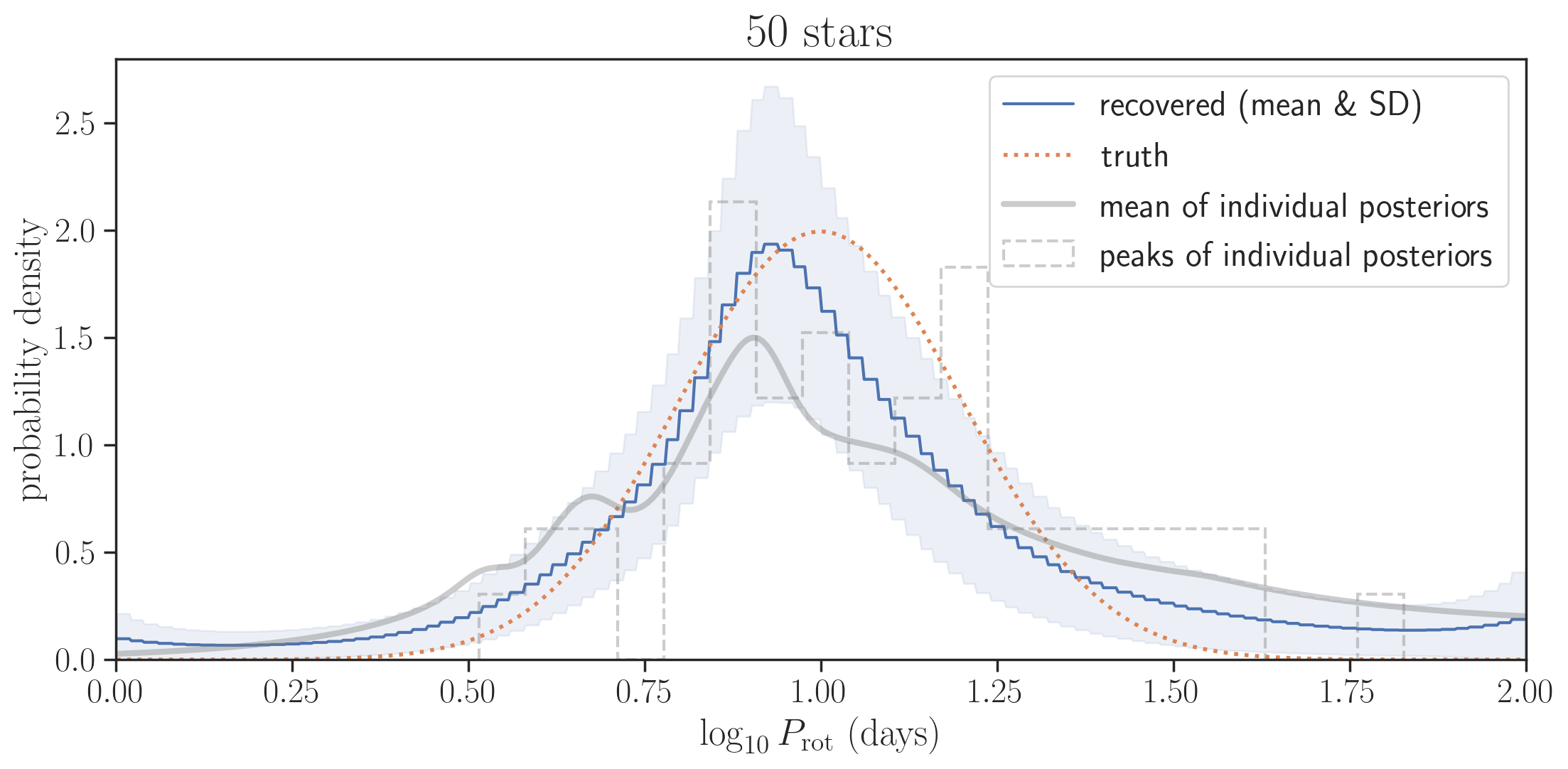}
	\includegraphics[width=0.81\columnwidth]{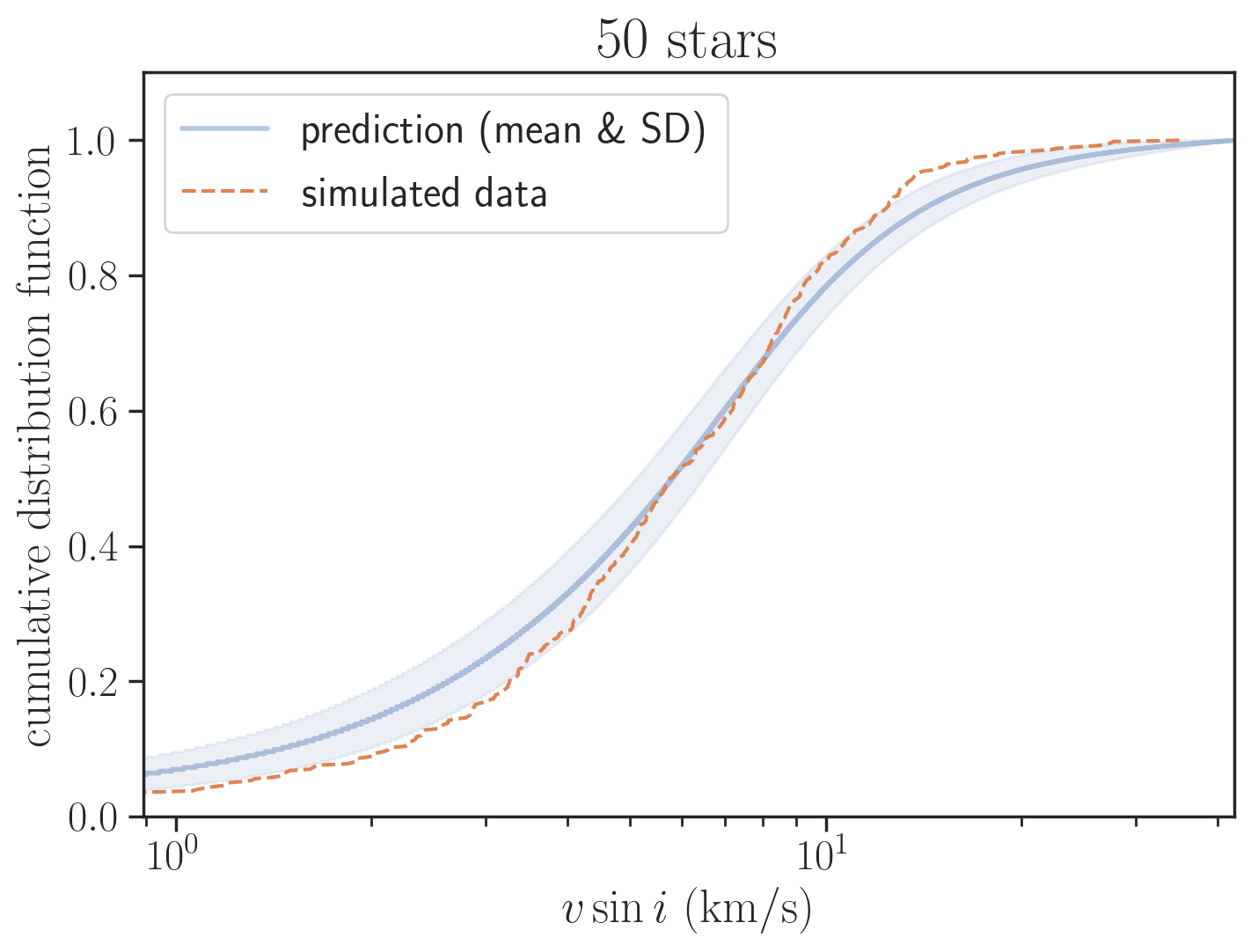}
	\includegraphics[width=1.26\columnwidth]{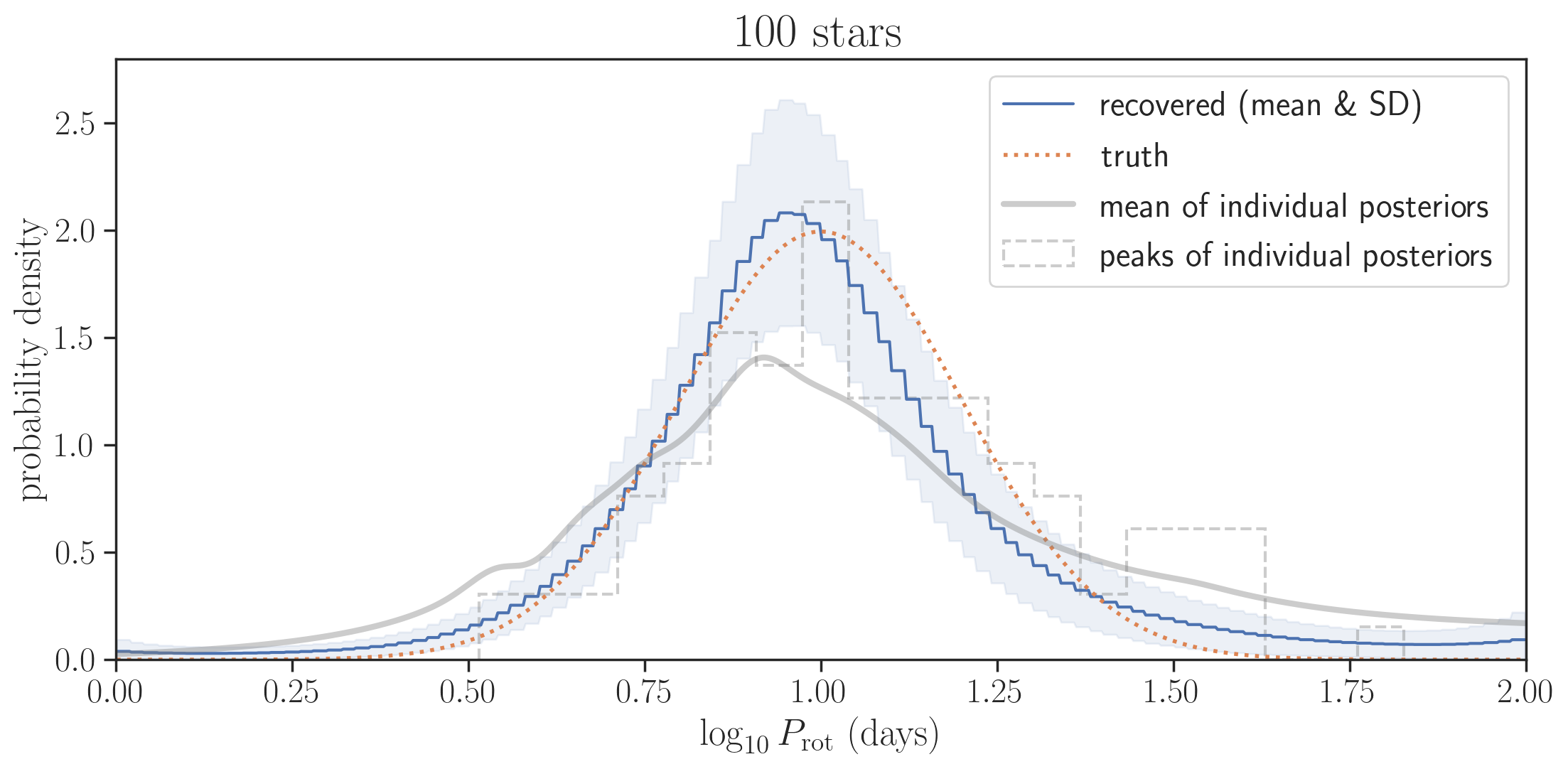}
	\includegraphics[width=0.81\columnwidth]{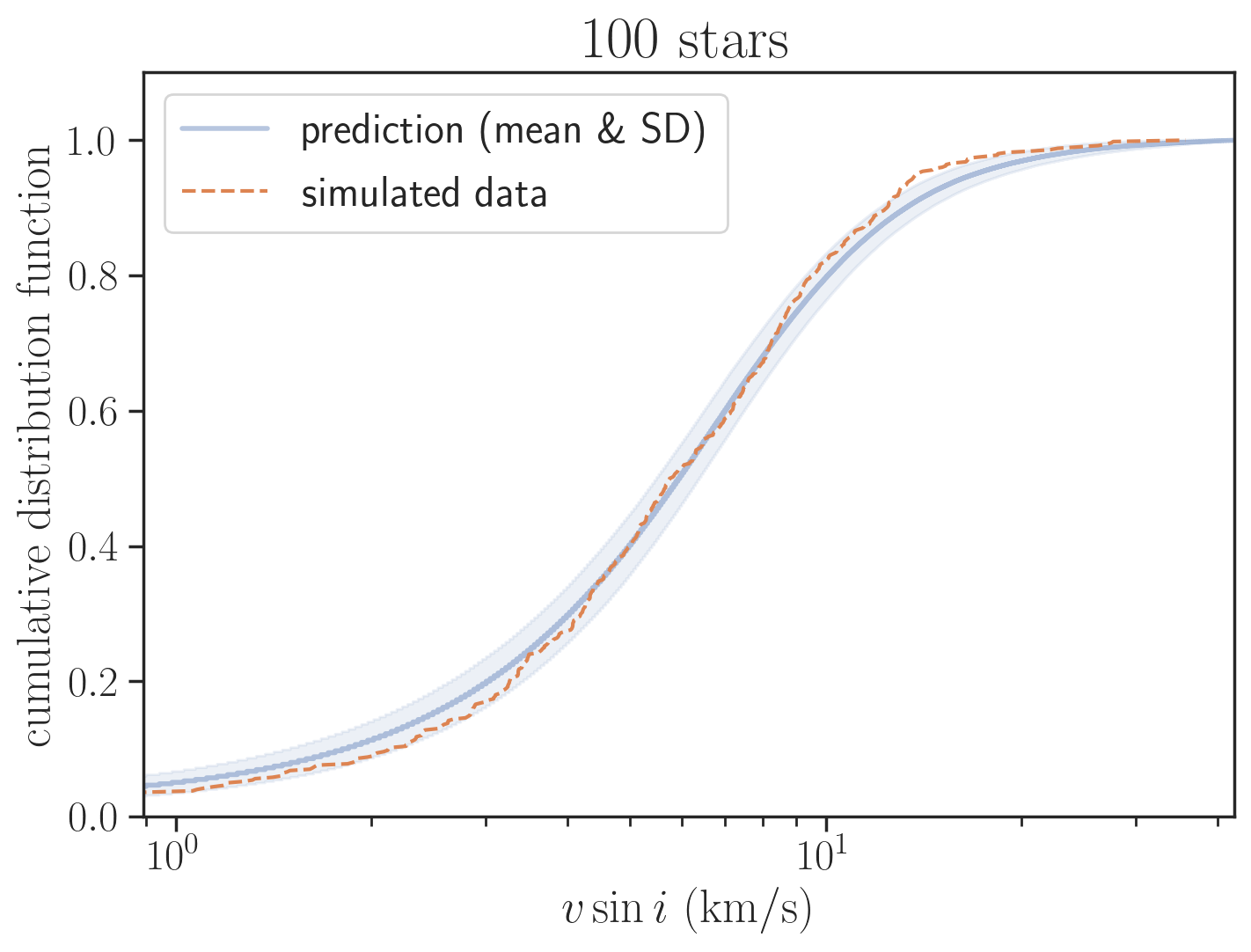}
	\includegraphics[width=1.26\columnwidth]{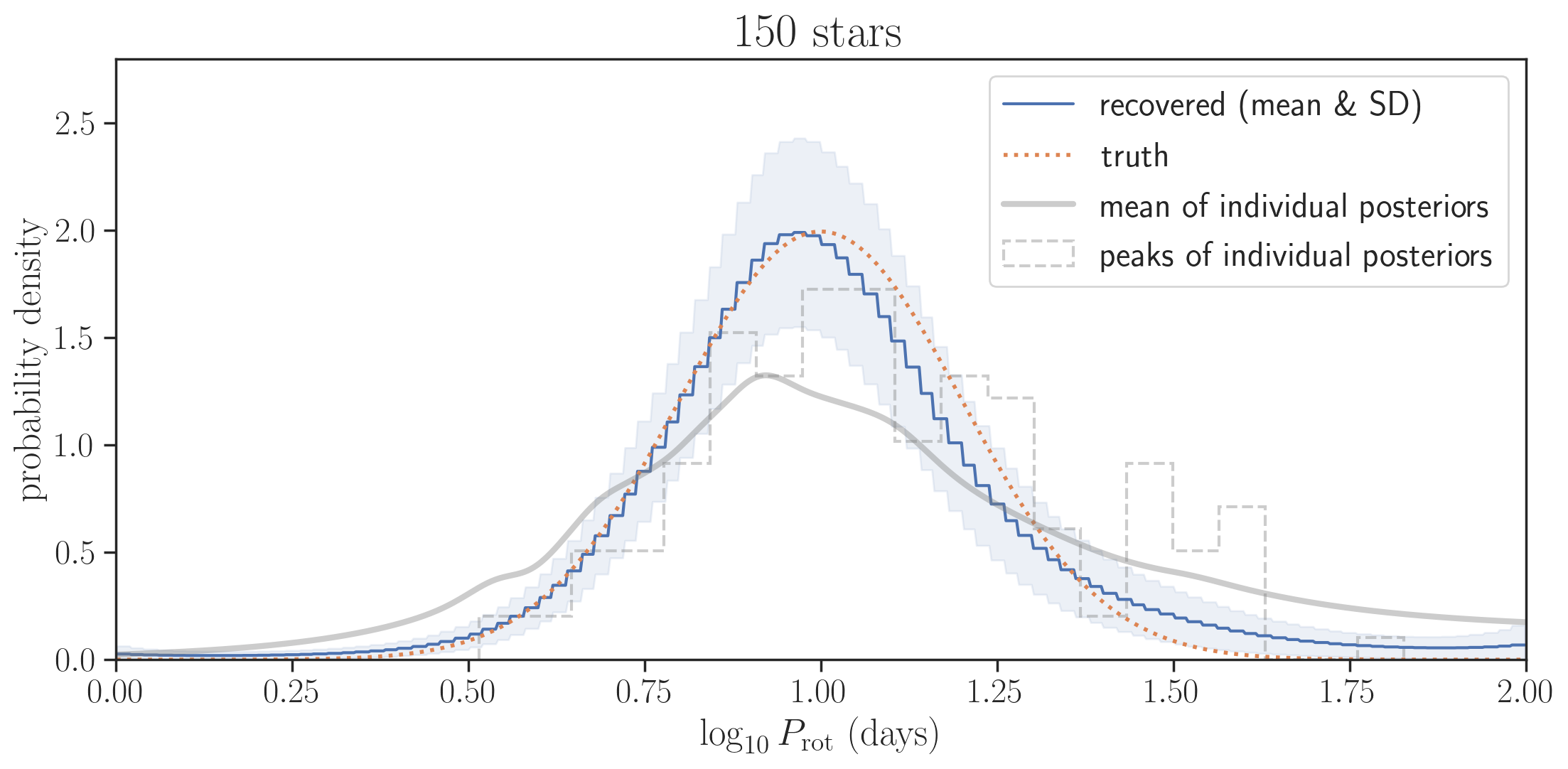}
	\includegraphics[width=0.81\columnwidth]{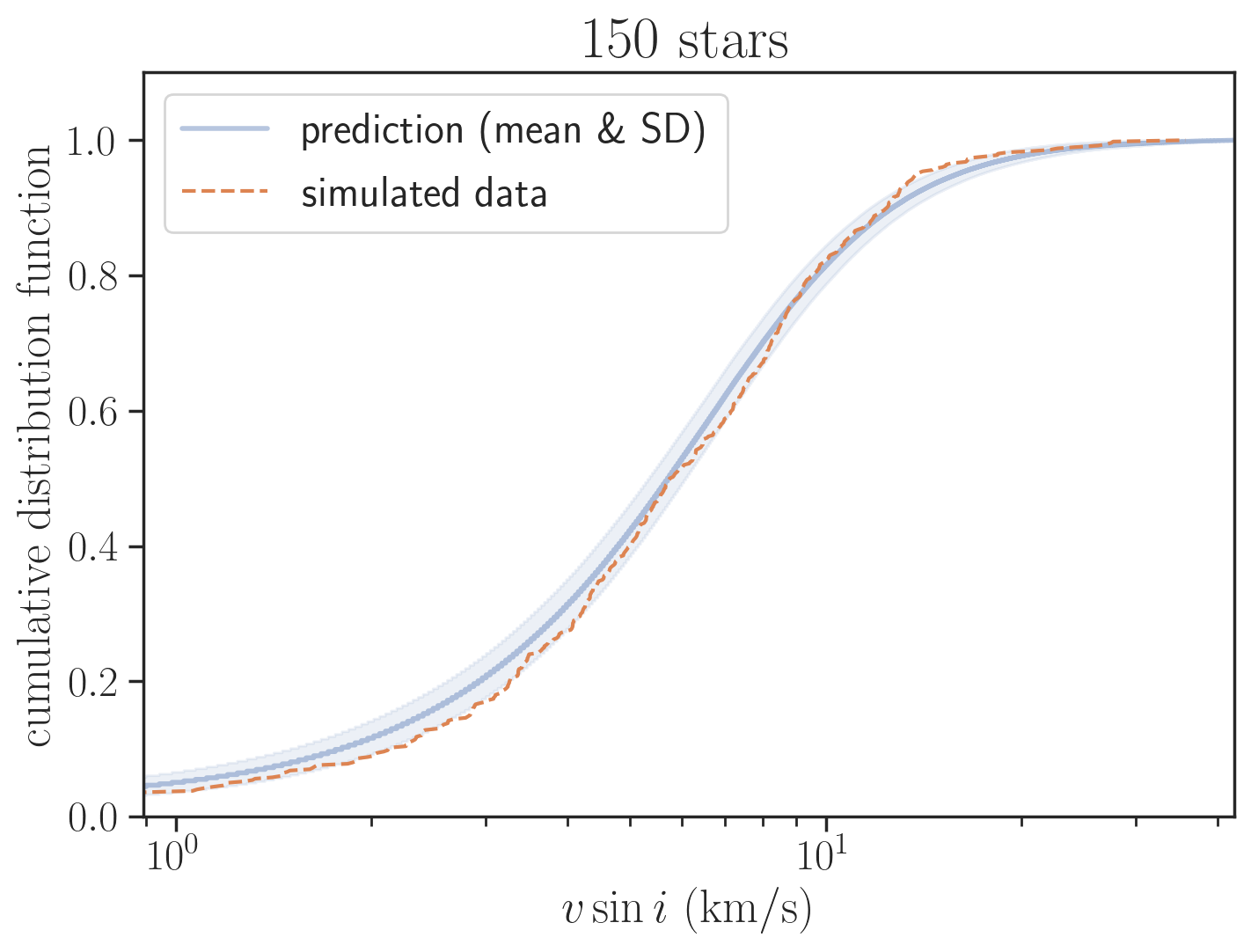}
    \caption{The recovery tests for a single log-normal input distribution (Section \ref{ssec:tests_single}): 50 stars (top), 100 stars (middle), and 150 stars (bottom). 
    In the left panels, the blue solid lines and shaded regions show the means and standard deviations (SD) of the inferred distribution. The orange dotted lines show the input distribution (truth). The thick gray lines show the means of the posterior PDFs for individual stars $p(x_j|D_j)$  assuming uniform priors, and the dashed gray histograms are created from the peaks of the individual posterior PDFs (i.e. MAP estimates); these are less optimal estimates of the true distribution as discussed in the text.
    In the right panels, the cumulative distribution function of the simulated $v\sin i$ (orange dashed line) is compared with that predicted from the inference (blue): periods are drawn from the inferred $\prot$ distribution, radii are drawn from the measured values considering their uncertainties, and $\cos i$ values are drawn uniformly from $(0,1)$.} 
    \label{fig:sim-lognorm}
\end{figure*}

\begin{table}
    \centering
    \caption{Predicted mean and standard deviation of $\logp$ (single log-normal input; Section \ref{ssec:tests_single}) for three different sample sizes. The quoted errors are the standard deviations calculated from the posterior samples of $\alpha$.}
    \label{tab:sim-lognorm}
    \begin{tabular}{l|cccc}
    \hline\hline
         & 50 stars & 100 stars & 150 stars  & truth \\
    \hline 
    mean & $1.01\pm0.05$ & $1.00\pm0.03$ & $1.01\pm0.03$ & 1\\
    standard deviation & $0.32\pm0.05$ & $0.27\pm0.03$ & $0.26\pm0.03$ & 0.2\\
    \hline
    \end{tabular}
\end{table}

\begin{figure*}
	\includegraphics[width=1.26\columnwidth]{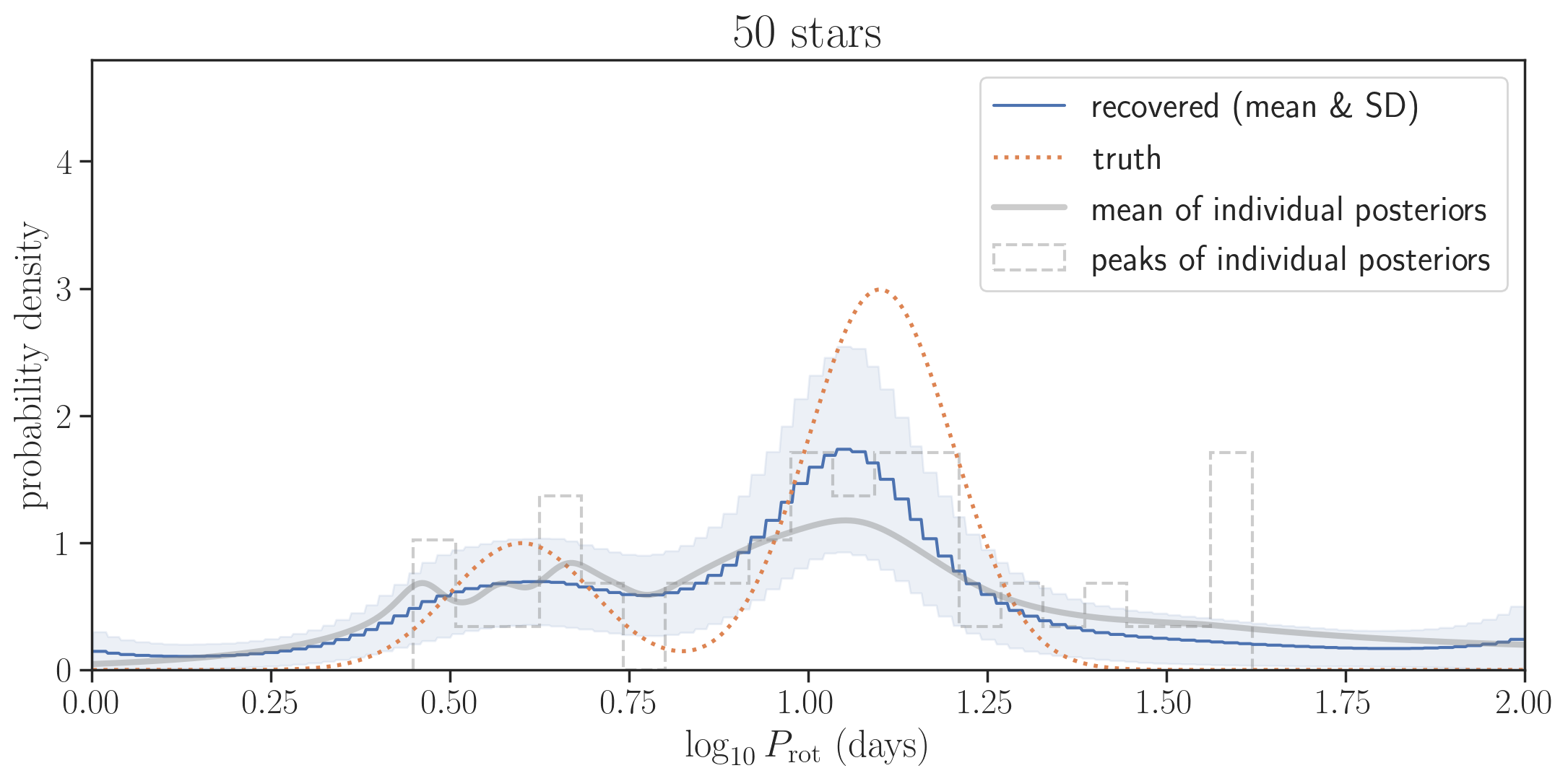}
	\includegraphics[width=0.81\columnwidth]{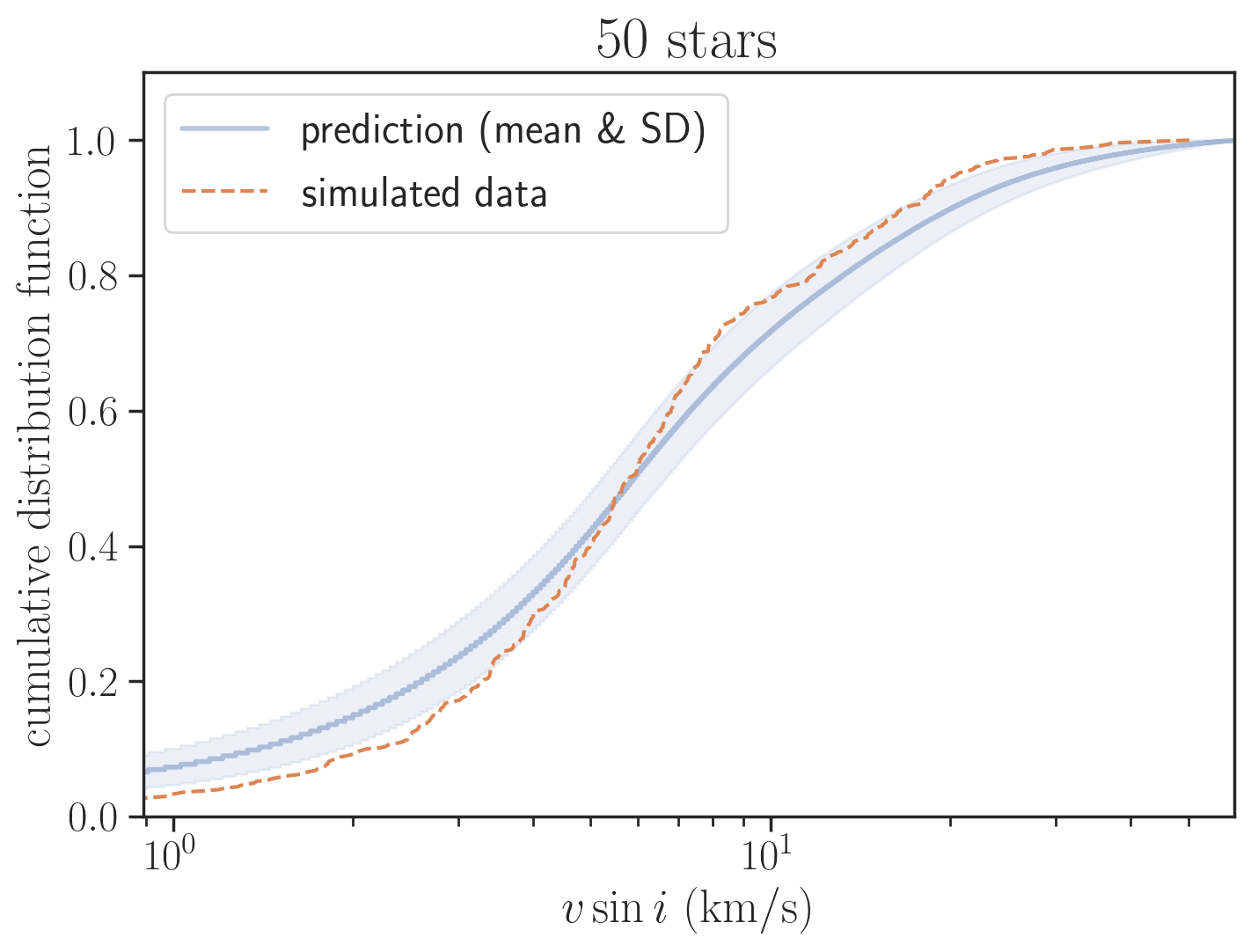}
	\includegraphics[width=1.26\columnwidth]{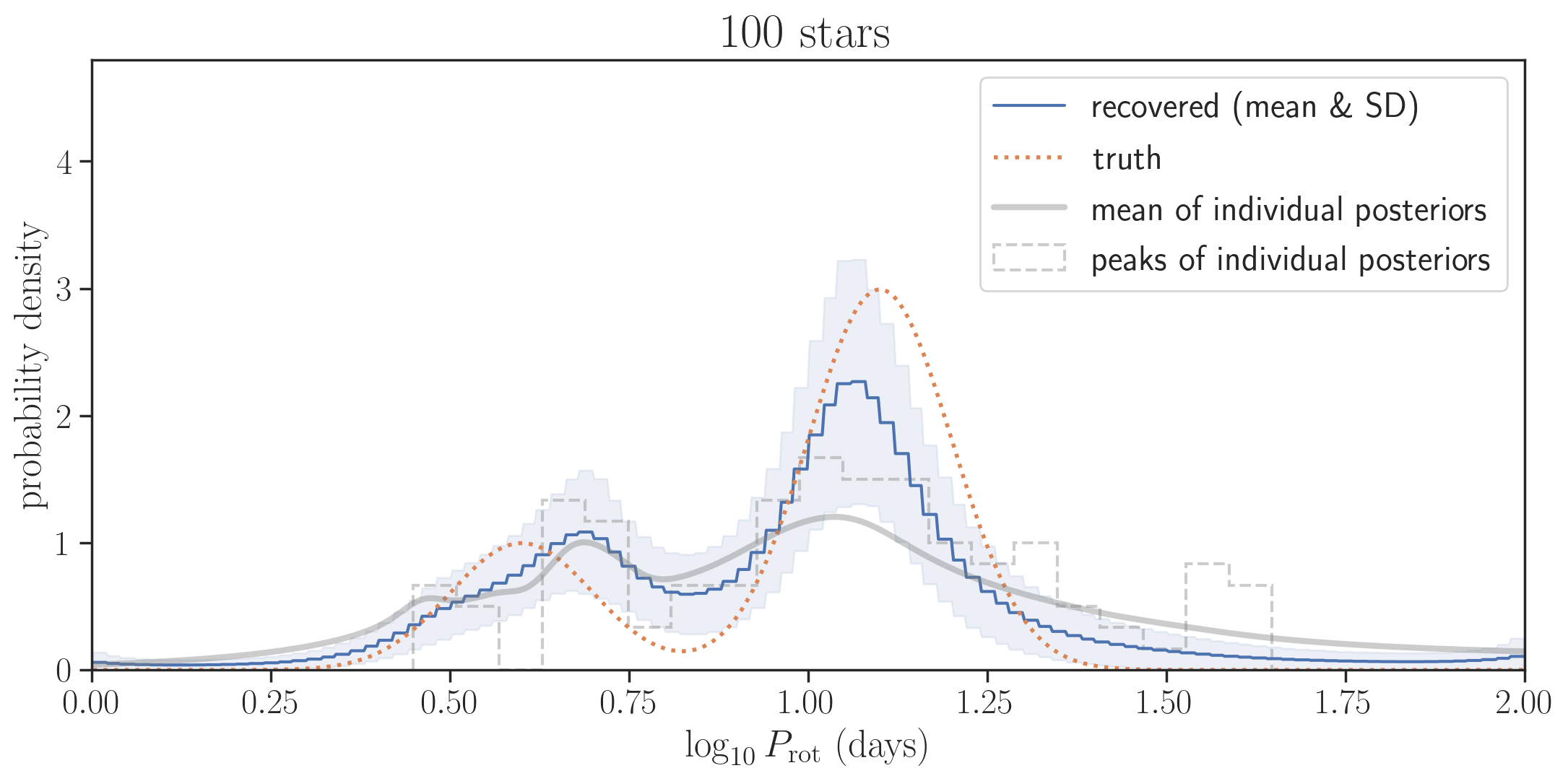}
	\includegraphics[width=0.81\columnwidth]{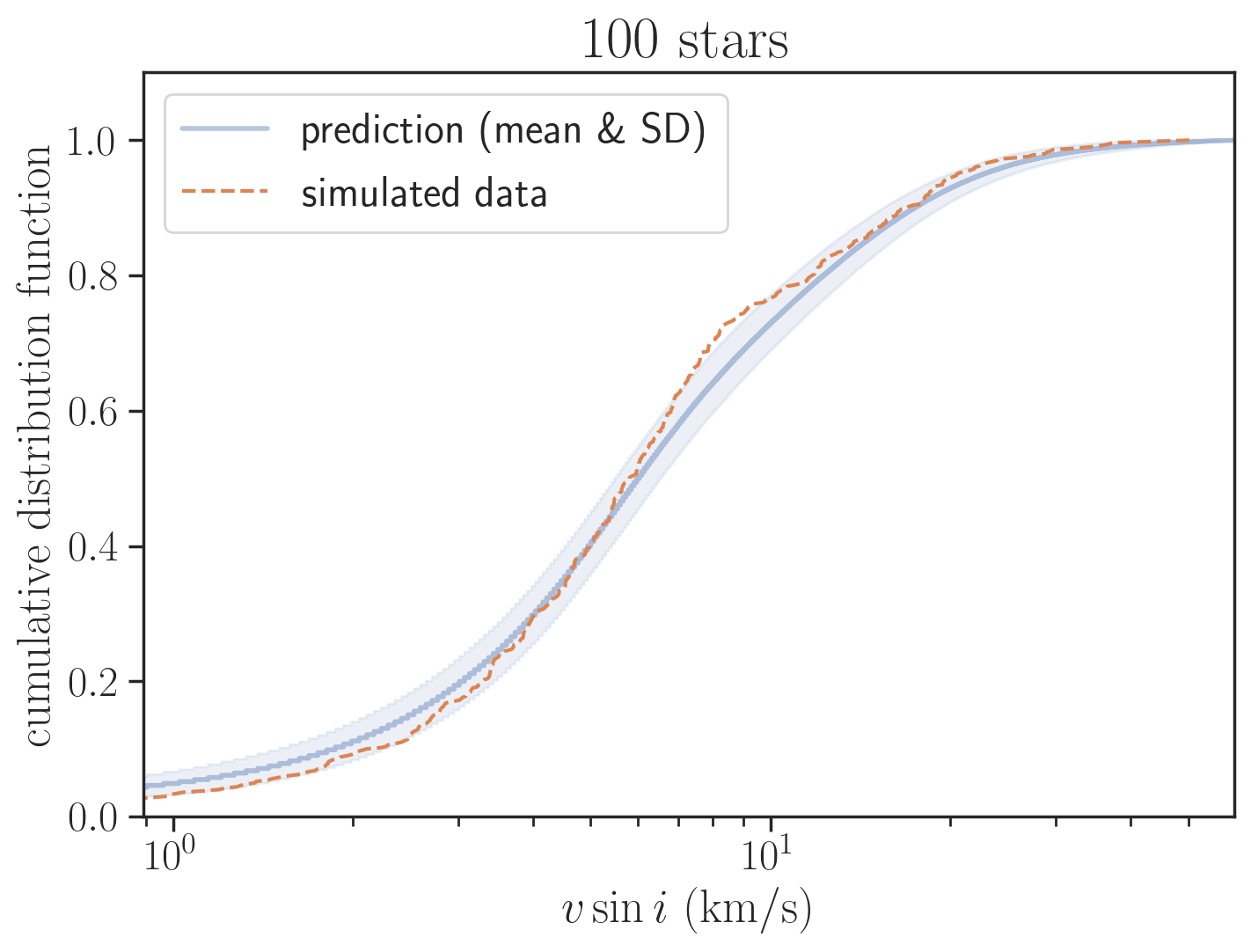}
	\includegraphics[width=1.26\columnwidth]{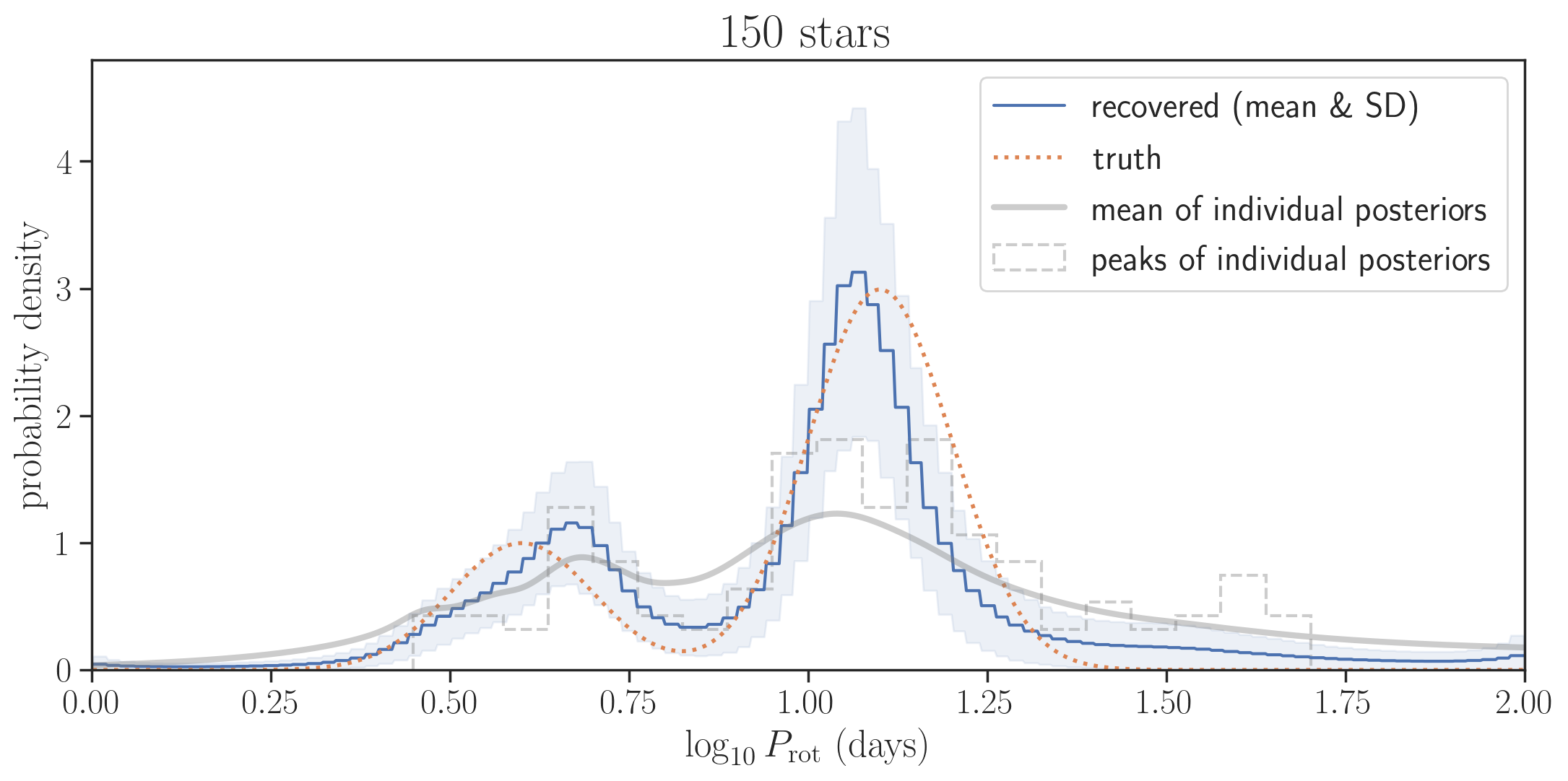}
	\includegraphics[width=0.81\columnwidth]{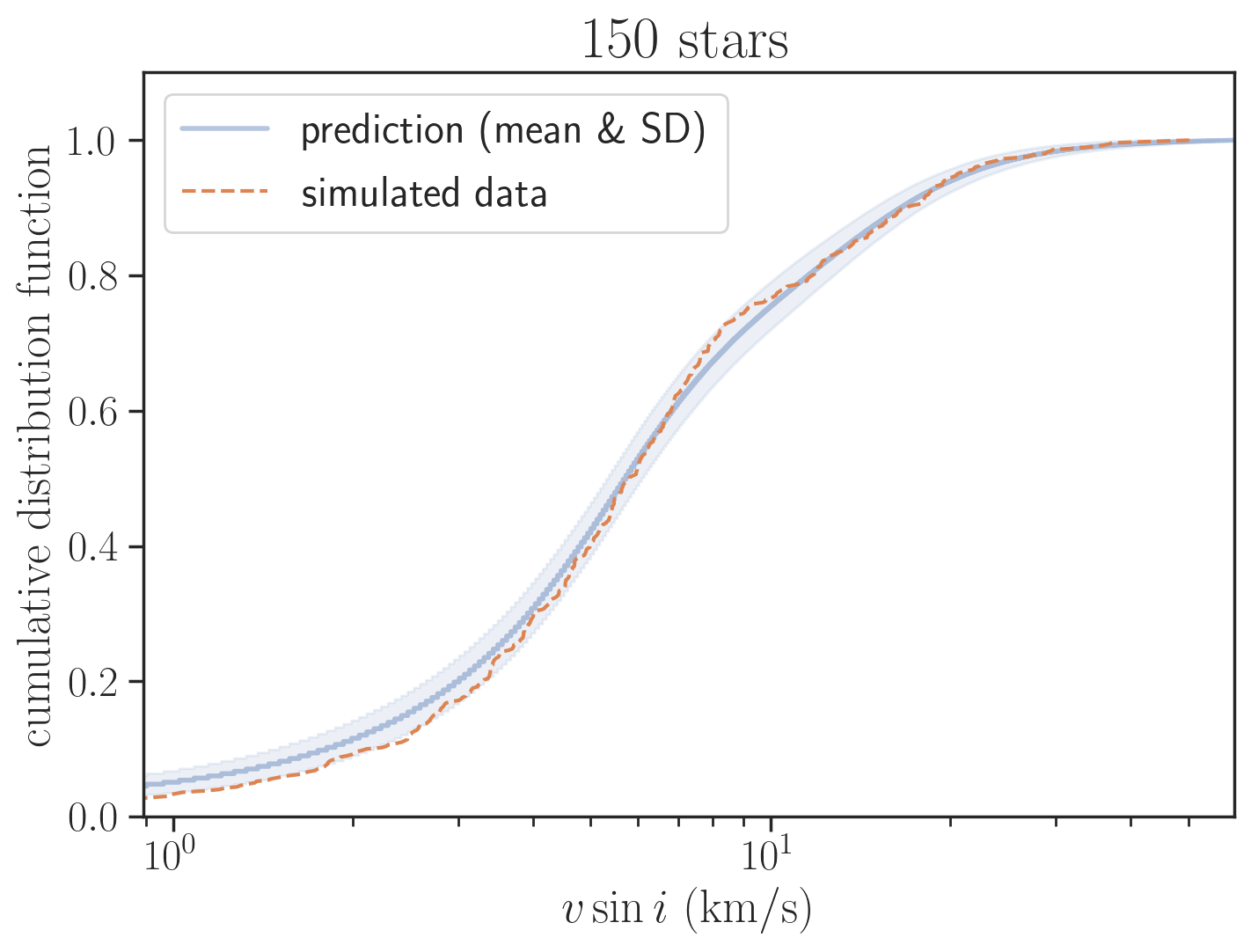}
    \caption{Same as Figure \ref{fig:sim-lognorm}, but for an input distribution that is a mixture of two log-normal distributions (Section \ref{ssec:tests_double}).}
    \label{fig:sim-2lognorm}
\end{figure*}

\begin{table}
    \centering
    \caption{Same as Table \ref{tab:sim-lognorm}, but for the double log-normal input distribution (Section \ref{ssec:tests_double}).}
    \label{tab:sim-2lognorm}
    \begin{tabular}{l|cccc}
    \hline\hline
         & 50 stars & 100 stars & 150 stars & truth \\
    \hline 
    mean &  $0.98\pm0.06$ & $0.96\pm0.03$ & $0.99\pm0.03$ & 0.97\\
    standard deviation & $0.38\pm0.05$ & $0.31\pm0.03$ & $0.30\pm0.03$ & 0.24\\
    \hline
    \end{tabular}
\end{table}

\begin{figure*}
	\includegraphics[width=1.26\columnwidth]{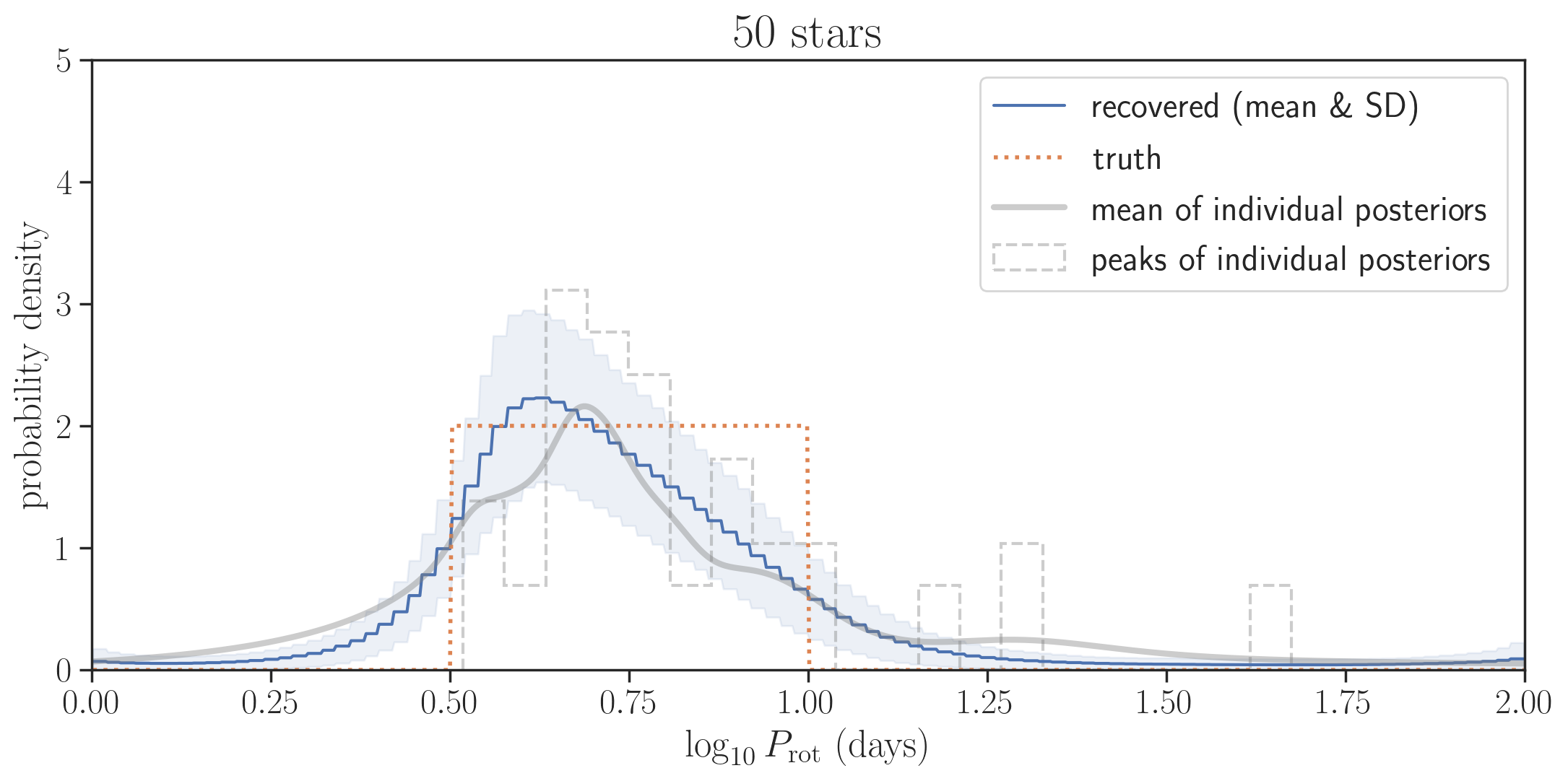}
	\includegraphics[width=0.81\columnwidth]{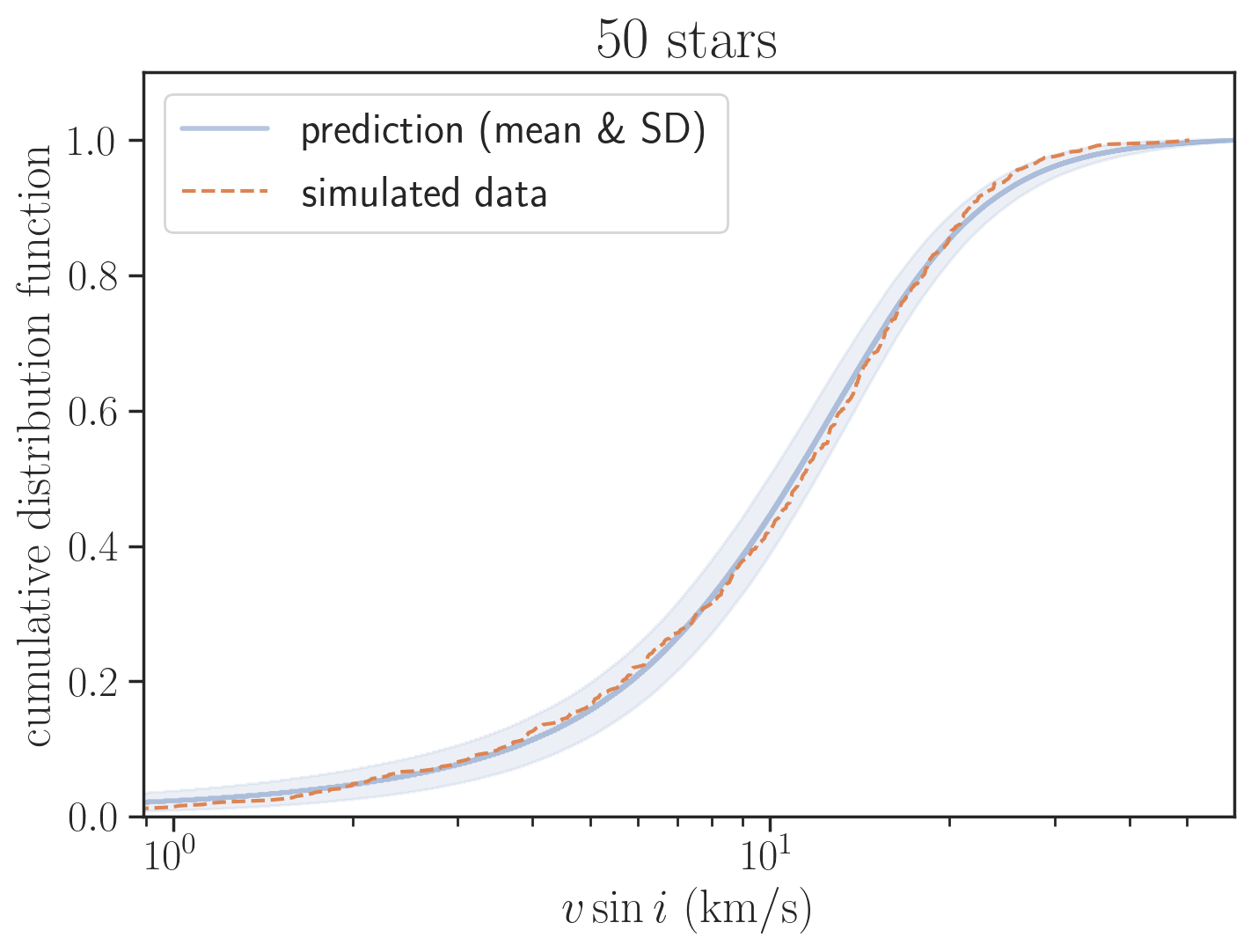}
	\includegraphics[width=1.26\columnwidth]{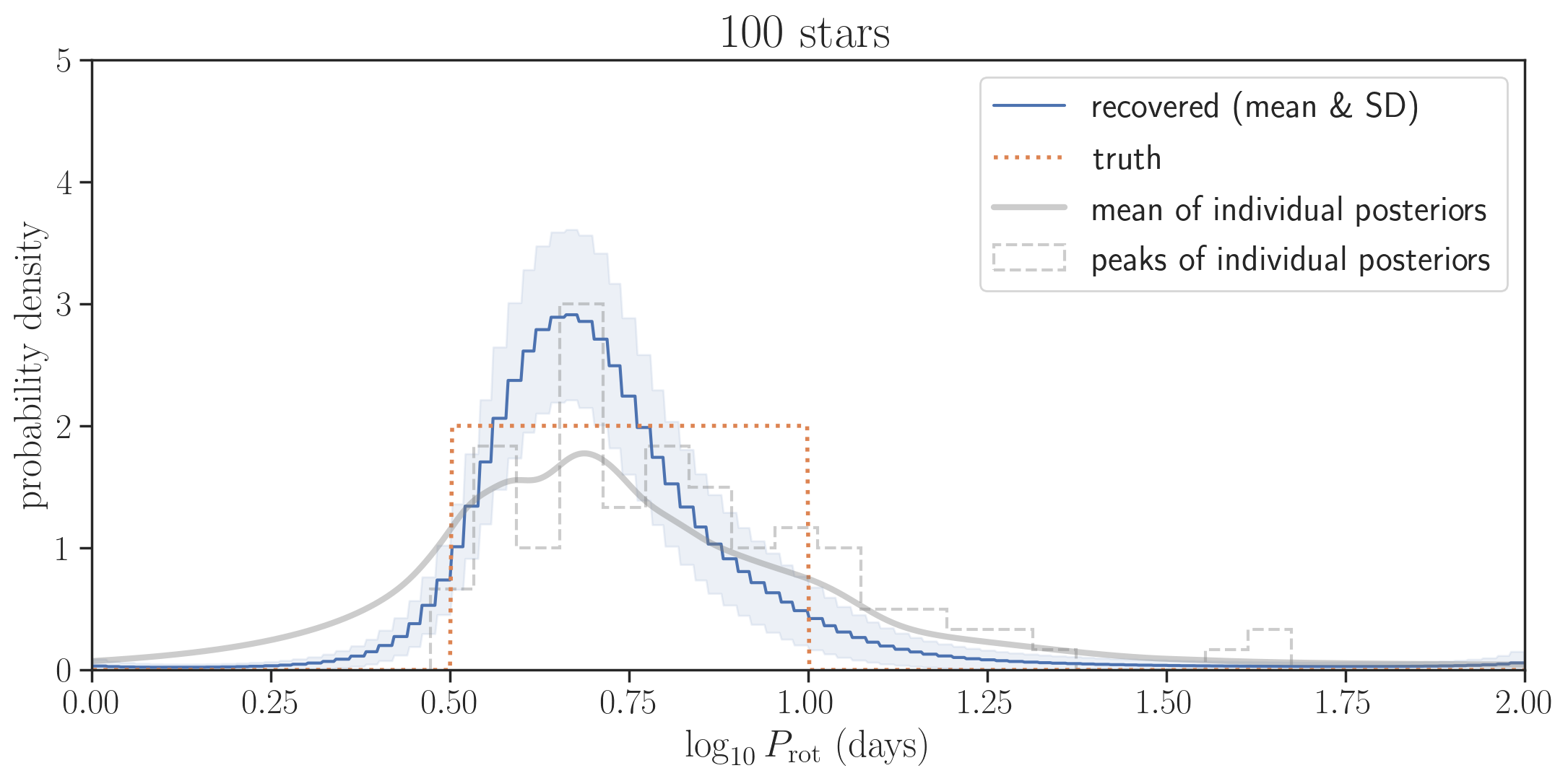}
	\includegraphics[width=0.81\columnwidth]{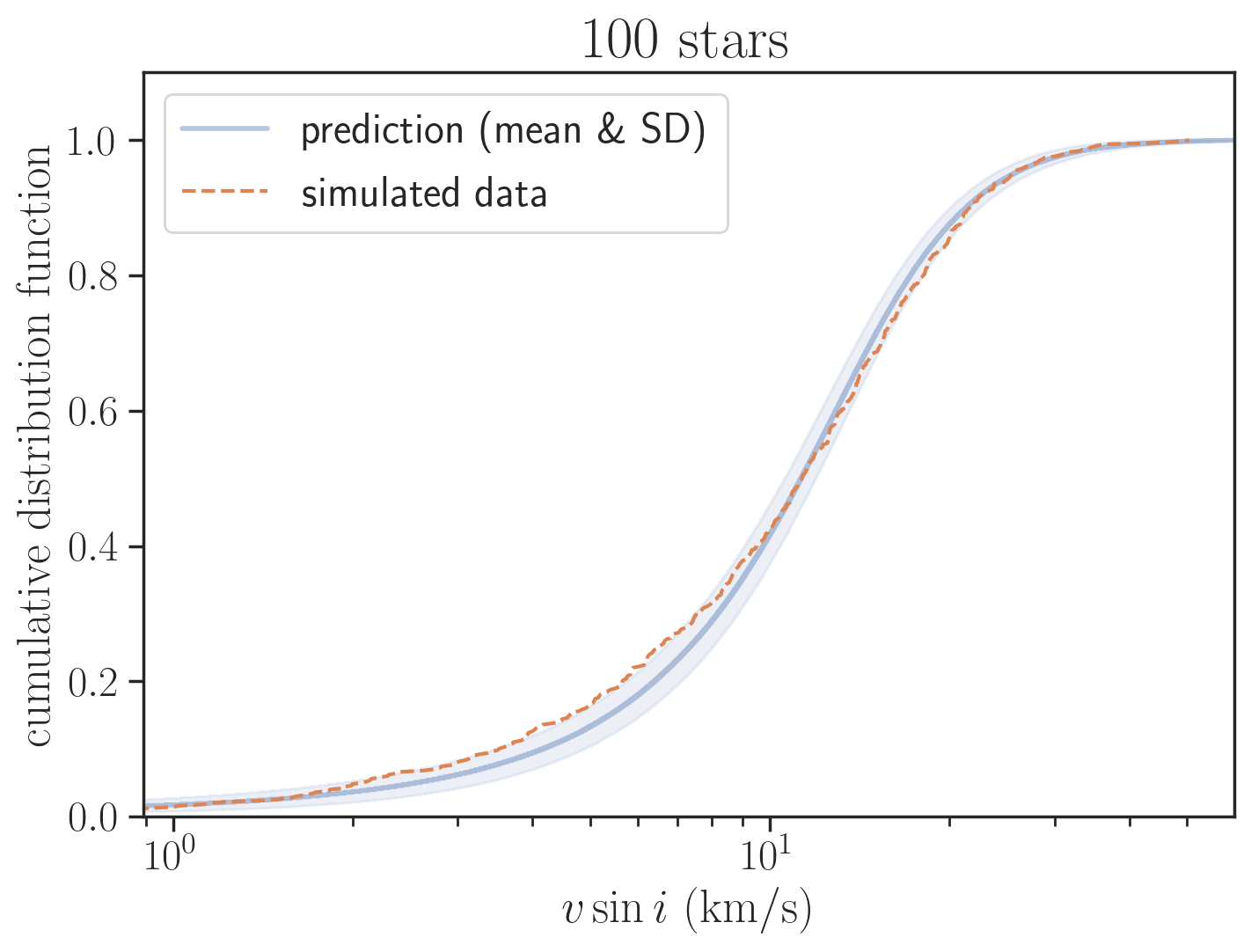}
    \includegraphics[width=1.26\columnwidth]{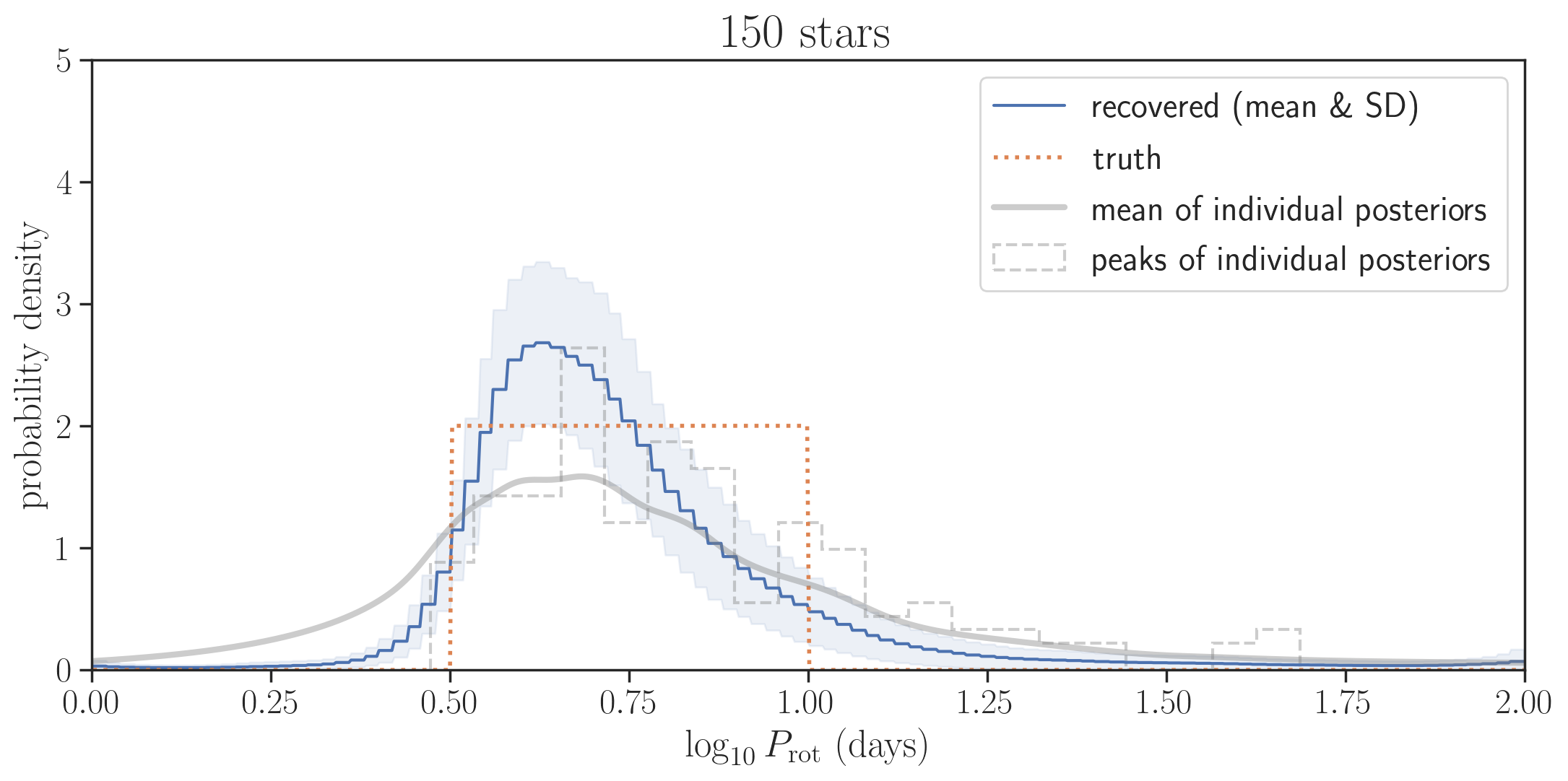}
	\includegraphics[width=0.81\columnwidth]{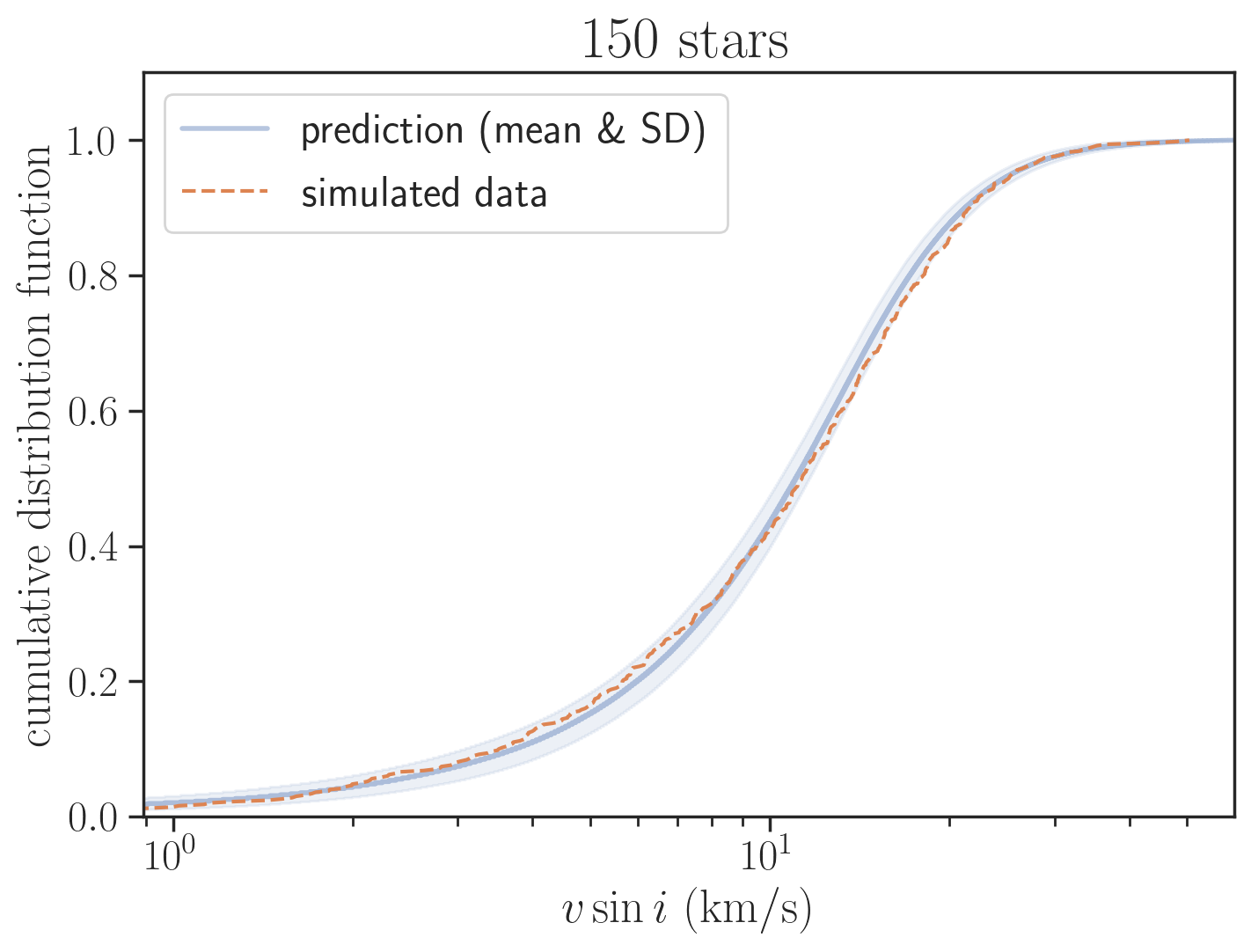}
    \caption{Same as Figure \ref{fig:sim-lognorm}, but for a log-uniform input distribution (Section \ref{ssec:tests_loguni}).}
    \label{fig:sim-loguni}
\end{figure*}

\begin{table}
    \centering
    \caption{Same as Table \ref{tab:sim-lognorm}, but for the log-uniform input distribution (Section \ref{ssec:tests_loguni}).}
    \label{tab:sim-loguni}
    \begin{tabular}{l|cccc}
    \hline\hline
         & 50 stars & 100 stars & 150 stars & truth \\
    \hline 
    mean &  $0.74\pm0.04$ & $0.73\pm0.03$ & $0.75\pm0.03$ & 0.75\\
    standard deviation & $0.26\pm0.06$ & $0.22\pm0.04$ & $0.24\pm0.04$ & 0.14\\
    \hline
    \end{tabular}
\end{table}

\section{Numerical Experiments using Simulated Data Sets}\label{sec:tests}

We apply the framework in Section \ref{sec:method} to simulated data sets and test in what situation the input $\prot$ distribution can (not) be successfully recovered. 
The data sets are created to mimic the actual measurements we will analyze in Section \ref{sec:kepler}.
We also show results of other sub-optimal estimates based on individual posteriors 
and discuss why they are not optimal.

In these experiments, we assume several functional forms for the PDF of $x \equiv \log_{10}(P/\mathrm{day})$ and draw ``true'' periods from this distribution. The true stellar radii are simulated by drawing from the measured radii of the \textit{Kepler} stars in our sample (Section \ref{sec:kepler}) with replacement and by perturbing their values using their respective errors. The true $\cos i$ value was sampled from a uniform distribution between $0$ and $1$ for each star. These are used to compute the true $v\sin i$ value of each star. 
Then the ``measured" values of $v\sin i$ and $R$ were simulated by adding zero-mean Gaussian random variables with the standard deviations of $1\,\mathrm{km/s}$ and $0.04R$
respectively to their true values, which approximate typical measurement uncertainties. The likelihood for $R$ was assumed to be a Gaussian centered on each simulated value with the $4\%$ relative width. The likelihood for $v\sin i$ was assumed to be a Gaussian centered on each simulated value with the width of $1\,\mathrm{km/s}$ when the measured value was $>1\,\mathrm{km/s}$; otherwise the likelihood was assumed to be flat and non-zero for $0\,\mathrm{km/s} < v\sin i < 2\,\mathrm{km/s}$ and zero otherwise (i.e., the measurement was considered to be an upper limit); these choices are based on the estimated precision of {\tt SpecMatch-Syn} \citep{2015PhDT........82P}.\footnote{In principle, one could use the likelihood functions directly derived from the data even in the case where it is appropriate to quote only ``upper limits.'' We do not do so in the present paper because the specific shapes of the likelihood functions in those cases are likely sensitive to systematics including macroturbulence (cf. Appendix \ref{sec:vmacro}) and would not be necessarily more accurate than the simple functional forms as adopted here.}
We applied the method in Section \ref{sec:method} to these simulated sets of likelihood functions $p(D_{u,j}|u)$ and $p(D_{R,j}|R)$.

\subsection{No Data}

We first tested the case where the likelihood $p(D_{u,j}, D_{R,j}|P_{\mathrm{rot},j})$ was independent from $P_{\mathrm{rot},j}$ for any $j$ (i.e., completely uninformative data). 
We found the mean posterior PDF for $x$ to be flat, as expected for uninformative data.

\subsection{Single Log-Normal Distribution} \label{ssec:tests_single}

Next, we assume that $x \equiv \logp$ follows a single normal distribution:
\begin{equation}
    f(x) = {1\over\sqrt{2\pi\sigma^2}}\exp\left[-{(x-\mu)^2 \over 2\sigma^2}\right],
\end{equation}
where $\mu=1$ and $\sigma=0.2$. 
In Figure \ref{fig:sim-lognorm}, the inferred distributions based on single realizations of 50, 100, and 150 samples (blue lines) are compared with the truth (orange dotted lines), and the estimated means/standard deviations of $x$ are compared with the truths in Table \ref{tab:sim-lognorm}. 
In this case, the prediction matches the input even for 50 stars within its standard deviation (blue shaded region), which we find to serve as a useful measure to gauge the deviation from the true distribution.
We also repeated the same analyses for 50 different realizations of 50, 100, and 150 stars, and confirmed that the mean of the mean predictions from the 50 realizations is consistent with the results shown here within one standard deviation. This means that these test results represent statistically typical outcomes. The same was found to be the case for the other two input period distributions discussed below.

In the left panels, we also show other estimates that can be more easily computed from individual posterior PDFs $p(x_j|D_j)$ assuming uniform priors but are not optimal: their mean (gray solid line) and peaks (gray dashed histogram).
The mean of the posteriors is always more dispersed than the input distribution, because they incorporate both measurement and $\sin i$ uncertainties.
Estimates using the posterior peaks (i.e., maximum a posteriori; MAP) show an excess of long-period stars because a certain fraction of stars always have $v\sin i$ much smaller than $v$ due to projection effects. For these reasons, these estimates do not approach the ground truth even as the sample size increases, while the hierarchical estimate does because it takes into account these effects in a consistent manner.

\subsection{Two Log-Normal Distributions} \label{ssec:tests_double}

Next, we try a mixture of two normal distributions:
\begin{equation}
    f(x) = {c_1\over\sqrt{2\pi\sigma_1^2}}\exp\left[-{(x-\mu_1)^2 \over 2\sigma_1^2}\right]
    + {c_2\over\sqrt{2\pi\sigma_2^2}}\exp\left[-{(x-\mu_2)^2 \over 2\sigma_2^2}\right],
\end{equation}
where we set $\mu_1=1.1$, $\mu_2=0.6$, $\sigma_1=\sigma_2=0.1$, $c_1=0.75$, and $c_2=0.25$. This functional form is inspired by the bimodality in the period distribution known for later-type \textit{Kepler} stars \citep{2013MNRAS.432.1203M, 2017ApJ...835...16D, 2018ApJ...868..151D, 2021ApJ...913...70G}, and is intended to check the sensitivity of our method to such a feature, if present.
The inferred distributions based on single realizations of 50, 100, and 150 samples are shown in Figure \ref{fig:sim-2lognorm}, and the estimated means/standard deviations are compared with the truths in Table \ref{tab:sim-2lognorm}. In this case, the bimodal feature is not totally clear with $50$ samples but becomes visible for $\gtrsim 100$ stars.

\subsection{Log-uniform Distribution: Failure Mode} \label{ssec:tests_loguni}

Our prior implements the (expected) smoothness of $f(x)$ and is unlikely to work well when the true distribution is not. To test this, we try a log-uniform distribution with sharp boundaries:
\begin{equation}
    f(x) = \begin{cases}
    2, \quad & 0.5<x<1\\
    0, \quad & \mathrm{otherwise}
    \end{cases}.
\end{equation}
The recovery results based on 50, 100, and 150 samples are shown in Figure \ref{fig:sim-loguni} and Table \ref{tab:sim-loguni}.
As expected, the input distribution is less well recovered than in the previous examples and tends to overestimate the dispersion significantly, although the mean values are still accurately inferred. Our current framework does not provide a means to identify such a situation, and this should be considered as a limitation of our method.

\section{Application to Late-F/early-G \textit{Kepler} Dwarfs} \label{sec:kepler}

Here we infer the $\prot$ distribution for a sample of late-F and early-G \textit{Kepler} stars ($5900\,\mathrm{K}\lesssim T_\mathrm{eff} \lesssim 6600\,\mathrm{K}$) from \citet{2021AJ....161...68L}. 
They compared the $v\sin i$ distribution of such hot stars with transiting planets against that of the control sample of similar stars without transiting planets, and searched for evidence of correlations between the inclinations of planetary orbits (nearly $90^\circ$ by construction) and those of stellar spins. 
Here we focus on the latter control stars that were selected without regard to the presence of transiting planets, for which spin orientations can be assumed to be isotropic.

\subsection{Stellar Sample and Parameters} \label{ssec:kepler_sample}

\citet{2021AJ....161...68L} used Keck/HIRES to obtain $R\approx 50,000$ spectra of 188 late-F/early-G \textit{Kepler} stars. We used their 2MASS IDs to 
find the corresponding \textit{Gaia} EDR3 sources \citep{2020arXiv201201533G} as well as other nearby sources within $10$~arcsec.
We removed 36 stars with bad astrometry and/or potential binary companions from the sample that satisfy either of the following criteria:
\begin{itemize}
    \item Keck/HIRES spectra show signs of secondary lines (19 stars)
    \item \textit{Gaia} parallax is not available, or {\tt parallax\_over\_error} is less than 10 (6 stars)
    \item multiple \textit{Gaia} sources exist within 2~arcsec (14 stars)
    \item \textit{Gaia} EDR3 metrics 
    suggest binarity \citep{2021A&A...649A...5F}: {\tt RUWE}$>1.4$ and {\tt harmonic\_amplitude}$>0.1$, or {\tt frac\_multi\_peak}$>2$ (25 stars)
\end{itemize}
This left us with 147 stars with apparently single-star spectra without obvious nearby companions. 
We also removed three 
stars with $\log g<3.9$ to obtain the final sample of 144 stars. The parameters of these sample stars are listed in Table \ref{tab:catalog}.

We need the radii and $v\sin i$ of these stars. For our fiducial analysis, we adopted the effective temperature $T_\mathrm{eff}$, surface gravity $\log g$, metallicity $\mathrm{[Fe/H]}$, and $v\sin i$ derived 
by applying the well-calibrated {\tt SpecMatch-Syn} pipeline \citep{2015PhDT........82P} to the Keck/HIRES spectra from \citet{2021AJ....161...68L}.\footnote{We found that $\log g$ from {\tt SpecMatch-Syn} may not be reliable for some of the stars, but confirmed that this potential issue has a negligible effect on the $v\sin i$ values (see Appendix \ref{sec:logg}).}
The stellar radii were then derived following the procedure in \citet{2018AJ....156..264F}, as implemented in the {\tt isoclassify} package \citep{2017ApJ...844..102H, 2020AJ....159..280B}. The dust map is from \citet{2019ApJ...887...93G}, and the parallax zero point for each star was corrected according to the recipe given by \citet{2020arXiv201201742L}. 
The median precision is $4\%$, which is around the floor of the systematic errors determined by the uncertainties of the fundamental temperature scale and stellar models \citep{2020arXiv201207957T}.
There are 10 stars in common with the asteroseismic sample of \citet{2021NatAs...5..707H}, whose spectroscopic $v\sin i$ range from $3.5$ to $11.8\,\mathrm{km/s}$. Excluding one star flagged above as a potential binary and two stars for which measurements were concluded to be less robust by \citet{2021NatAs...5..707H}, the $v\sin i$ difference (spectroscopic minus seismic) is found to be $-0.3\pm0.9\,\mathrm{km/s}$ (mean and standard deviation). For the stellar radii we find the difference to be $0.02\pm0.02\,R_\odot$.

We also estimated masses and ages of our sample stars by fitting the MIST models \citep{2011ApJS..192....3P, 2013ApJS..208....4P, 2015ApJS..220...15P, 2016ApJS..222....8D, 2016ApJ...823..102C} to the spectroscopic $T_\mathrm{eff}$, $\mathrm{[Fe/H]}$, 2MASS $K_s$ magnitude, and \textit{Gaia} EDR3 parallax with errors inflated by $15\%$ \citep{2021MNRAS.tmp..394E}, adopting the extinction from \citet{2018MNRAS.478..651G}.
We adopted the medians and standard deviations of the posterior samples obtained with the {\tt isochrones} package \citep{2015ascl.soft03010M} as summary statistics. Again we found an excellent agreement with asteroseismology: for the same stars in common with \citet{2021NatAs...5..707H} as discussed above, the age and mass differences (mean and standard deviation) were $0.0\pm1.0\,\mathrm{Gyr}$ and $0.0\pm0.08\,M_\odot$, respectively, where the seismic parameters were from \citet{2015MNRAS.452.2127S, 2017ApJ...835..173S}.
The $\log g$ cut left 137 stars instead of 144 when adopting those isochrone-based parameters, but we found that the result was insensitive to this difference. 

\begin{table}
    \centering
    \caption{Information on the 144 stars used for the main analysis. Only first five entries are shown. Full table is available online. Other parameters are also available from \url{https://github.com/kemasuda/prot_from_vsini}.}
    \label{tab:catalog}
    \begin{tabular}{cccc}
    \hline\hline
    KIC & $T_{\rm eff}$ (K) & $v\sin i$ (km/s) & radius ($R_\odot$) \\
    \hline 
2158850 & 6038 & 2.7 & $1.15\pm0.05$\\
2998253 & 6291 & 9.2 & $1.45\pm0.06$\\
3338777 & 5993 & 6.9 & $2.37\pm0.10$\\
3831297 & 6111 & 3.6 & $1.13\pm0.04$\\
3936993 & 6176 & 7.9 & $1.85\pm0.07$\\
$\cdots$ & $\cdots$ & $\cdots$ & $\cdots$\\
    \hline
    \end{tabular}
\end{table}

\begin{figure}
	\includegraphics[width=1\columnwidth]{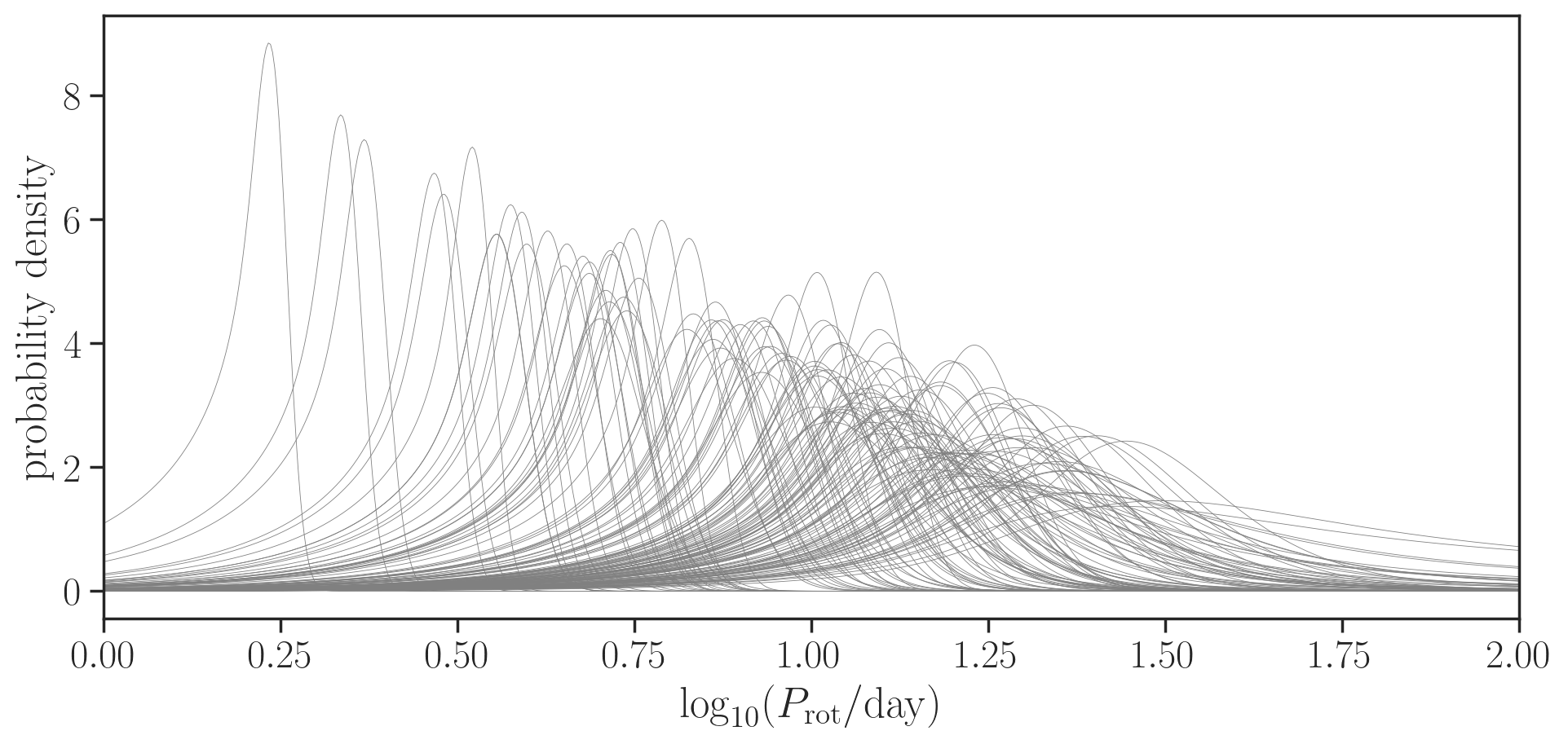}
    \caption{Posterior PDFs of $\logp$ for all 144 stars in the sample assuming the prior uniform in $\logp$. This choice was made here only to illustrate the form of $p(D_j|\logp)$, and the actual analysis was performed using the hierarchical prior as described in Section \ref{ssec:method_framework}.}
    \label{fig:all-pdfs}
\end{figure}

\begin{figure*}
	\includegraphics[width=1.26\columnwidth]{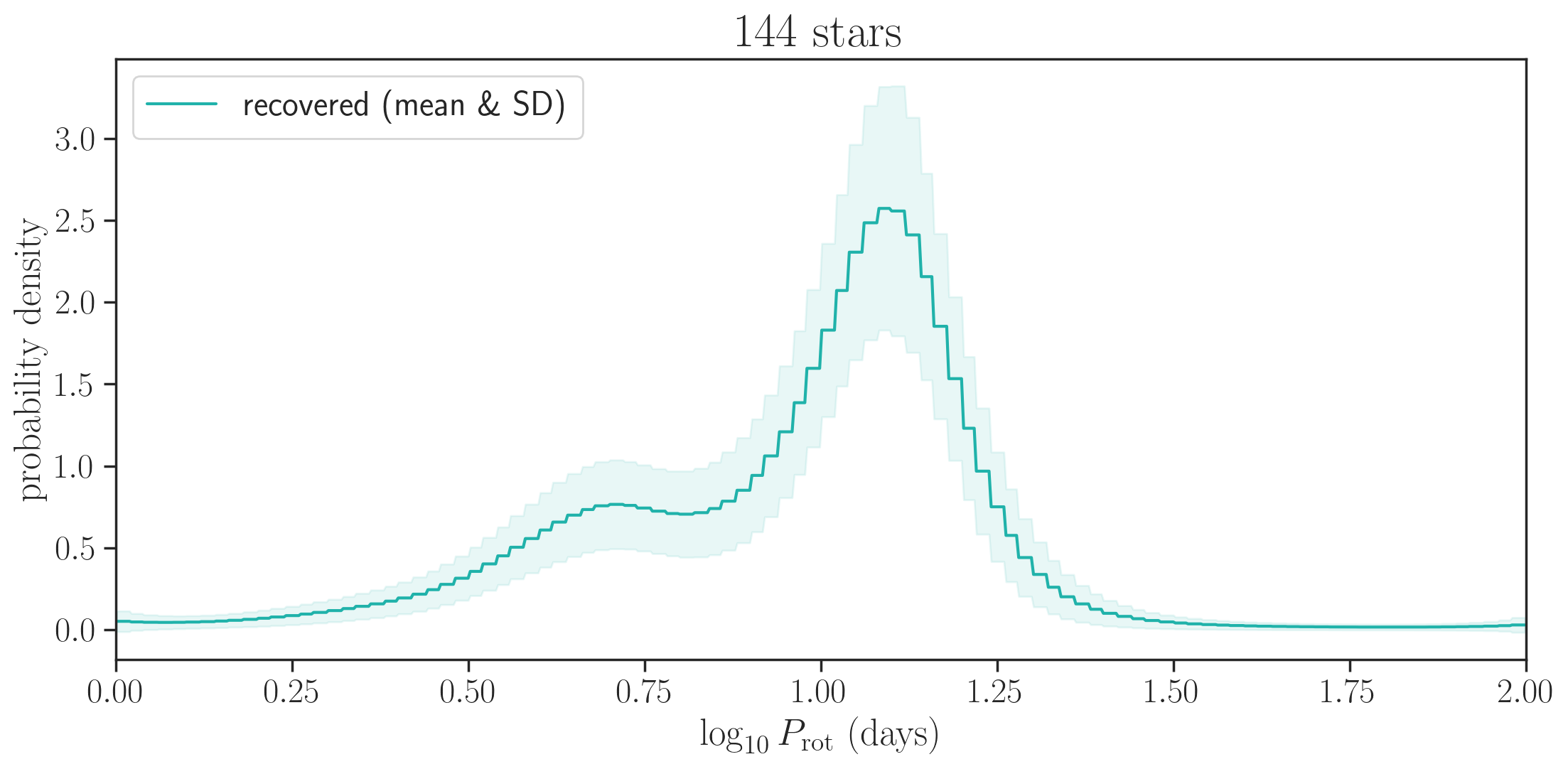}
	\includegraphics[width=0.81\columnwidth]{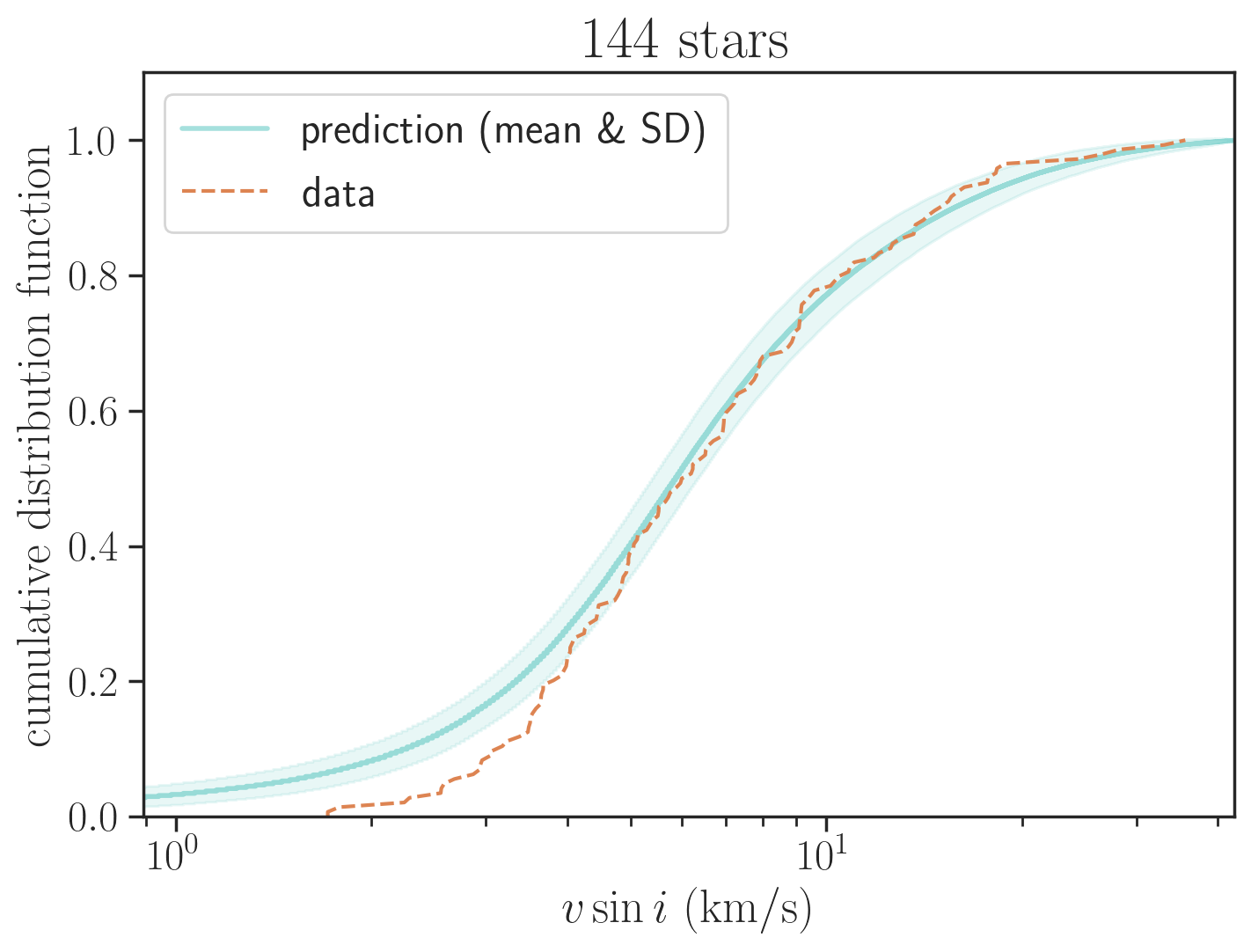}
    \caption{(Left) The $\log_{10} P$ distribution inferred for the whole sample under the assumptions stated in the first paragraph of Section \ref{ssec:kepler_results}. Here the solid line and shaded region show the mean and standard deviation of the recovered distribution, respectively. (Right) 
    The $v\sin i$ distribution predicted from the period distribution in the left panel (green line and shaded region), compared against the data (orange dashed line).
    The model predicts more stars with $v\sin i \lesssim 3\,\mathrm{km/s}$ than observed, suggesting that the spectroscopic $v\sin i$ is overestimated in this range. Figure \ref{fig:upper3} presents the result that takes into account this potential bias.
    }
    \label{fig:all}
\end{figure*}

\begin{figure*}
	\includegraphics[width=1.26\columnwidth]{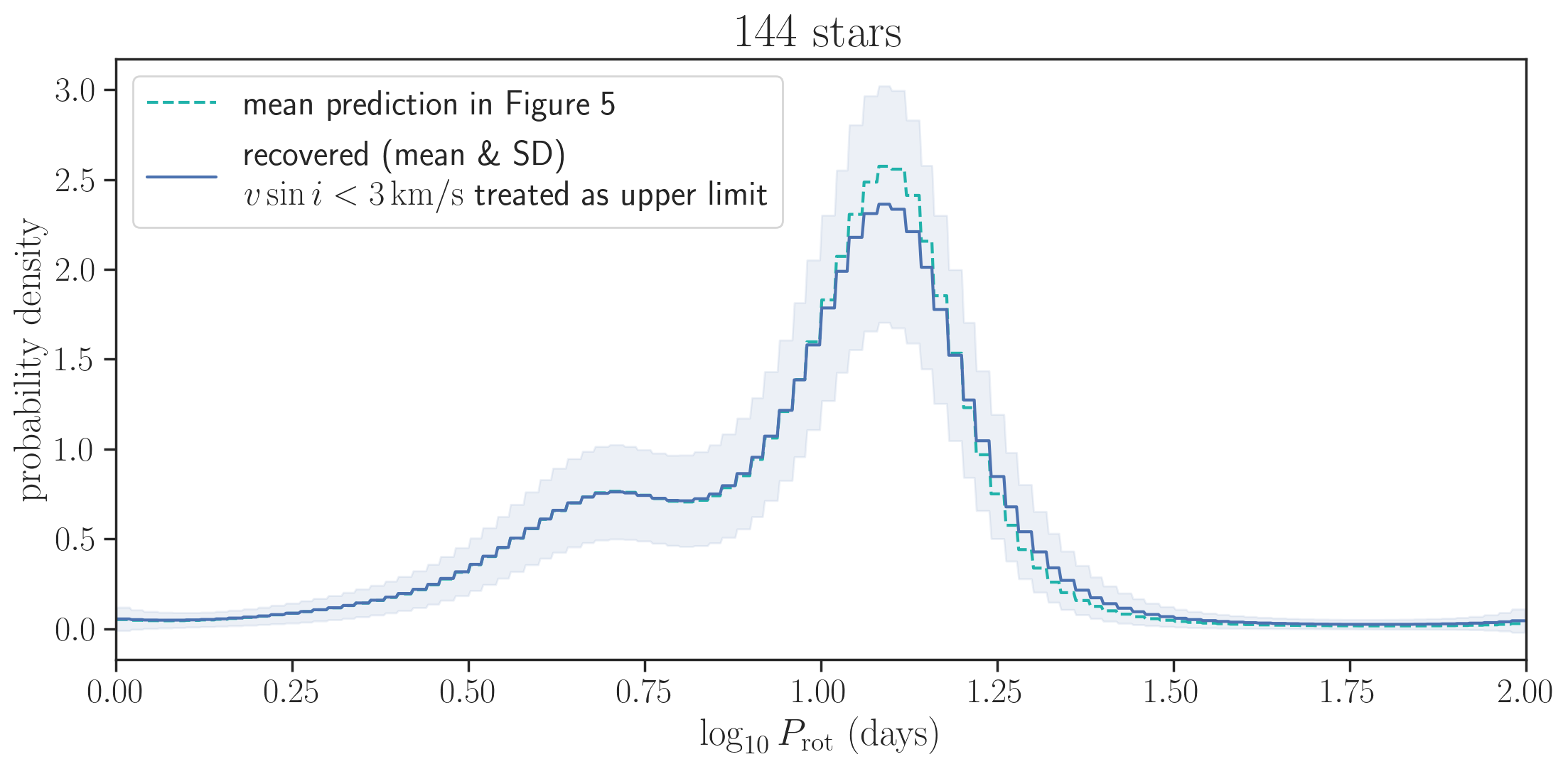}
	\includegraphics[width=0.81\columnwidth]{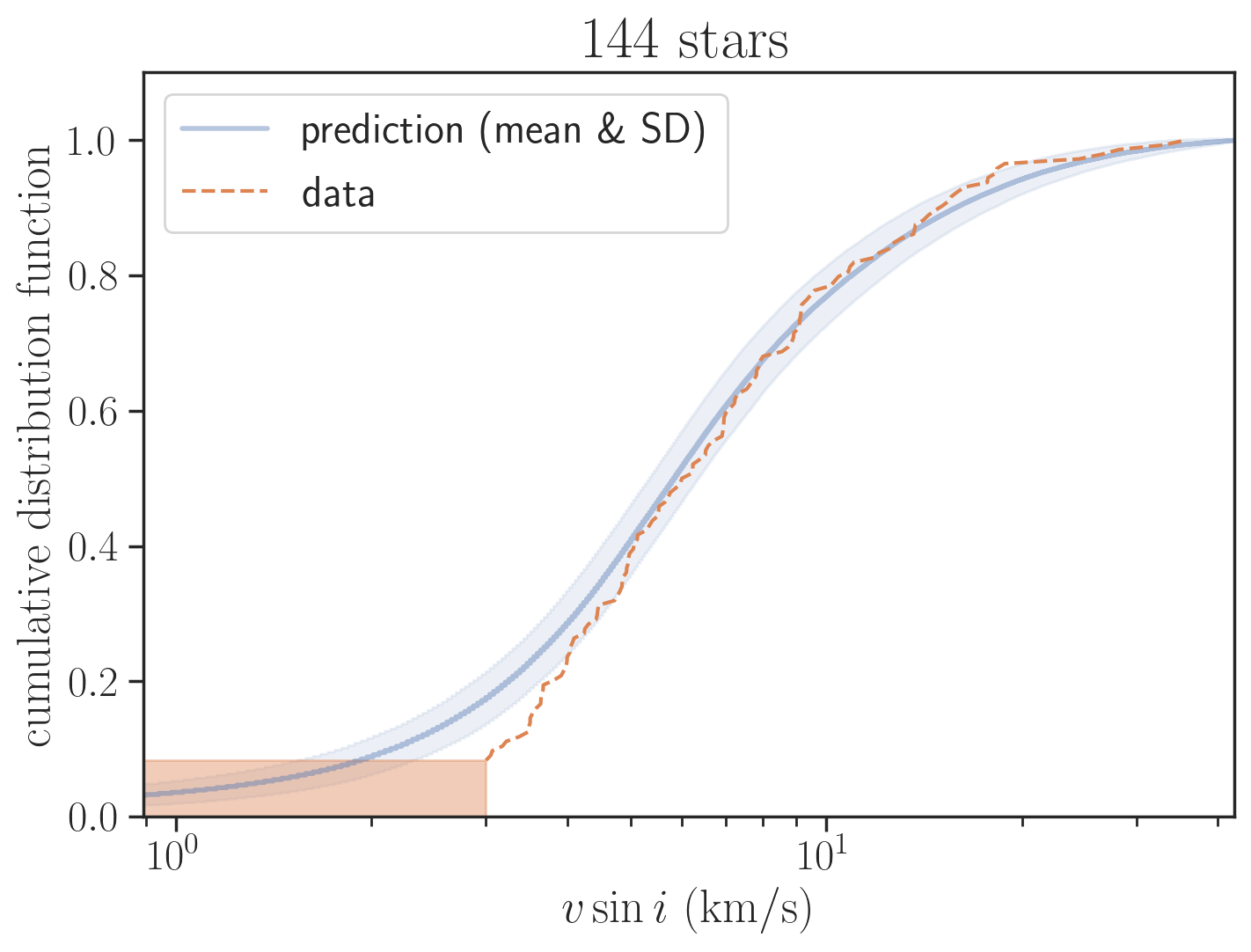}
    \caption{
    (Left) The $\log_{10} P$ distribution inferred for the whole sample treating $v\sin i<3\,\mathrm{km/s}$ to be upper limits (see Section \ref{ssec:kepler_results} for details). The solid blue line and shaded region show the mean and standard deviation of the recovered distribution, respectively. The green dashed line shows the mean prediction in Figure \ref{fig:all}. The good agreement between the two indicates that the recovered $\prot$ distribution is insensitive to potential systematic errors at low $v\sin i$.
    (Right) 
    The $v\sin i$ distribution predicted from the period distribution in the left panel (blue line and shaded region), compared against the data (orange dashed line). The orange box indicates that the measurements yielding $v\sin i<3\,\mathrm{km/s}$ are  treated as upper limits.
    }
    \label{fig:upper3}
\end{figure*}

\subsection{
Rotation Period Distribution of \textit{Kepler} F/G Dwarfs} \label{ssec:kepler_results}

Figure \ref{fig:all-pdfs} displays the posterior PDFs $p(x_j|D_j)$ computed from the marginal likelihood $p(D_j|x_j)$ in Equation~\ref{eq:margL} for all the 144 stars in the sample, adopting a uniform prior PDF for $x_j$ (this prior is adopted only in this figure for illustration).
Here we adopted Gaussian likelihoods for $R$ based on the {\tt isoclassify} outputs; for $v\sin i$, we adopted Gaussian likelihoods centered on the {\tt SpecMatch-Syn} measurements with $1\,\mathrm{km/s}$ width for $v\sin i>1\,\mathrm{km/s}$, and a likelihood flat over $[0\,\mathrm{km/s},2\,\mathrm{km/s}]$ for $v\sin i<1\,\mathrm{km/s}$ (although this latter choice will be modified later).
Figure \ref{fig:all} (left) shows the mean (green line; Equation \ref{eq:meanposterior}) and standard deviation (shaded region) of the inferred period distribution. The mean and standard deviation of $\logp$ are estimated to be $0.96\pm0.02$ ($\approx 9\,\mathrm{days}$) and $0.26\pm0.02$ ($\approx 2\,\mathrm{days}$), respectively. Figure \ref{fig:all} (right) compares the $v\sin i$ distribution predicted from the inferred period distribution against the data, where $\prot$ was drawn from the recovered distribution, $R$ was drawn from the same Gaussian as adopted for the likelihood, and $\cos i$ was drawn from the uniform distribution. Unlike in the analyses of the simulated data in Section \ref{sec:tests}, the model predicts too many stars with $v\sin i\lesssim3\,\mathrm{km/s}$; equivalently, there is a lack of small $v\sin i$ stars that should be present if the spins are isotropically oriented. This suggests that spectroscopic $v\sin i$ is overestimated in this range, as was also suggested in previous comparisons with asteroseismology \citep{2018MNRAS.479..391K, 2021NatAs...5..707H}.
We do not believe that the discrepancy is due to the lack of flexibility of our period model, because such a discrepancy was not seen in the tests using simulated data (Section \ref{sec:tests}) and because the discrepancy remained even when the bin heights were inferred without any informative prior (in which case the resulting distribution was not smooth).

Motivated by the above tension, as well as a good agreement with the seismic values found for $v\sin i\gtrsim 3.5\,\mathrm{km/s}$ (Section \ref{ssec:kepler_sample}), we performed another round of analysis treating all the measurements with $v\sin i<3\,\mathrm{km/s}$ to be the same upper limits: for those stars the likelihood was assumed to be flat and non-zero for $0\,\mathrm{km/s} < v\sin i < 3\,\mathrm{km/s}$ and $0$ otherwise, regardless of the assigned $v\sin i$ values. The results are shown in Figure \ref{fig:upper3}. We find that the inferred period distribution remains unaffected, because it is largely determined by the other $\approx 90\%$ stars with $v\sin i>3\,\mathrm{km/s}$. The mean and standard deviation of $\logp$ are estimated to be $0.97\pm0.03$ and $0.27\pm0.02$, respectively. Thus we conclude that a potential systematic error at low $v\sin i$ does not affect the result significantly. To be conservative, though, we adopt the results based on this latter model that treats $v\sin i<3\,\mathrm{km/s}$ to be upper limits (i.e., those measurements were considered less informative).

\begin{figure*}
	\includegraphics[width=2\columnwidth]{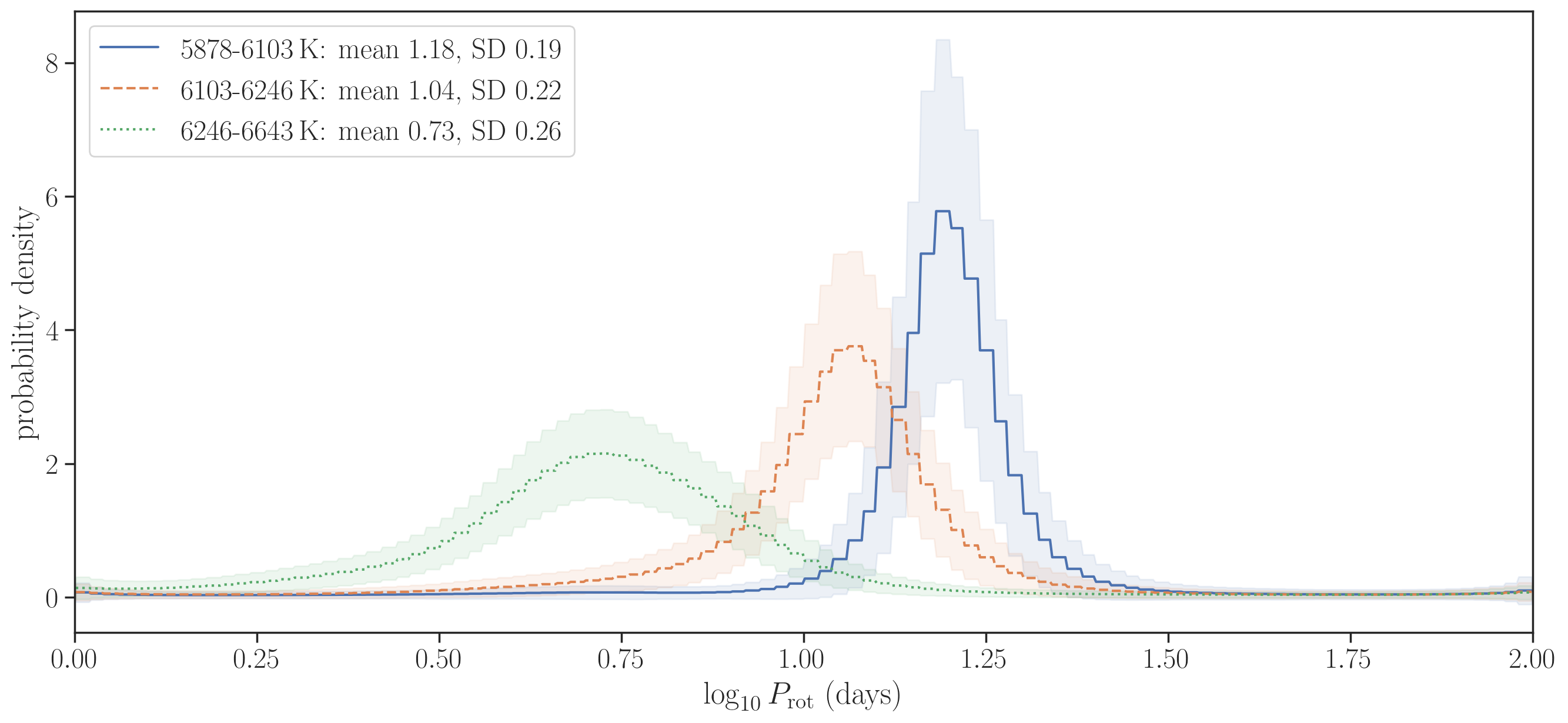}
    \caption{The $\log_{10} P$ distributions inferred for the subsamples with different effective temperatures shown in the legend. Also shown are the mean and standard deviation (SD) of $\logp$ for the mean distribution in each temperature range.}
    \label{fig:upper3_temp}
\end{figure*}

\begin{figure*}
	\includegraphics[width=2\columnwidth]{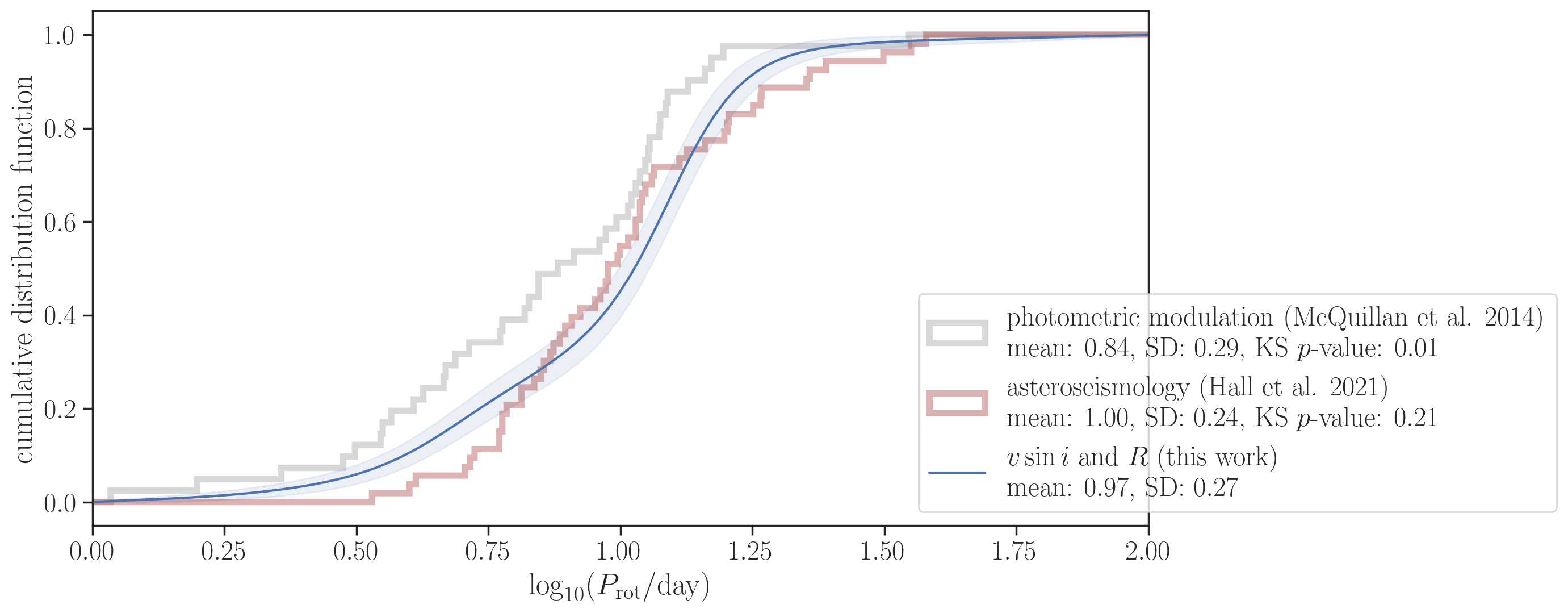}
    \caption{The rotation period distribution and its standard deviation inferred from $v\sin i$ and $R$ (blue solid line and shaded region) compared with periods measured using photometric modulation (gray) for the 41 stars in our sample by \citet{2014ApJS..211...24M}.
    The red histogram shows the asteroseismic rotation periods derived by \citet{2021NatAs...5..707H}
    for 53 stars with $5900\,\mathrm{K}<T_\mathrm{eff}<6600\,\mathrm{K}$, $\log g>3.9$, and Flag $0$ (indicating robust measurements);
    they are mostly different stars from our sample (seven stars in common) but share similar stellar properties.
    For the latter two, the $p$-values of the KS tests against the mean distribution inferred from $v\sin i$ and $R$ are also shown.}
    \label{fig:prot_comparison}
	\includegraphics[width=2\columnwidth]{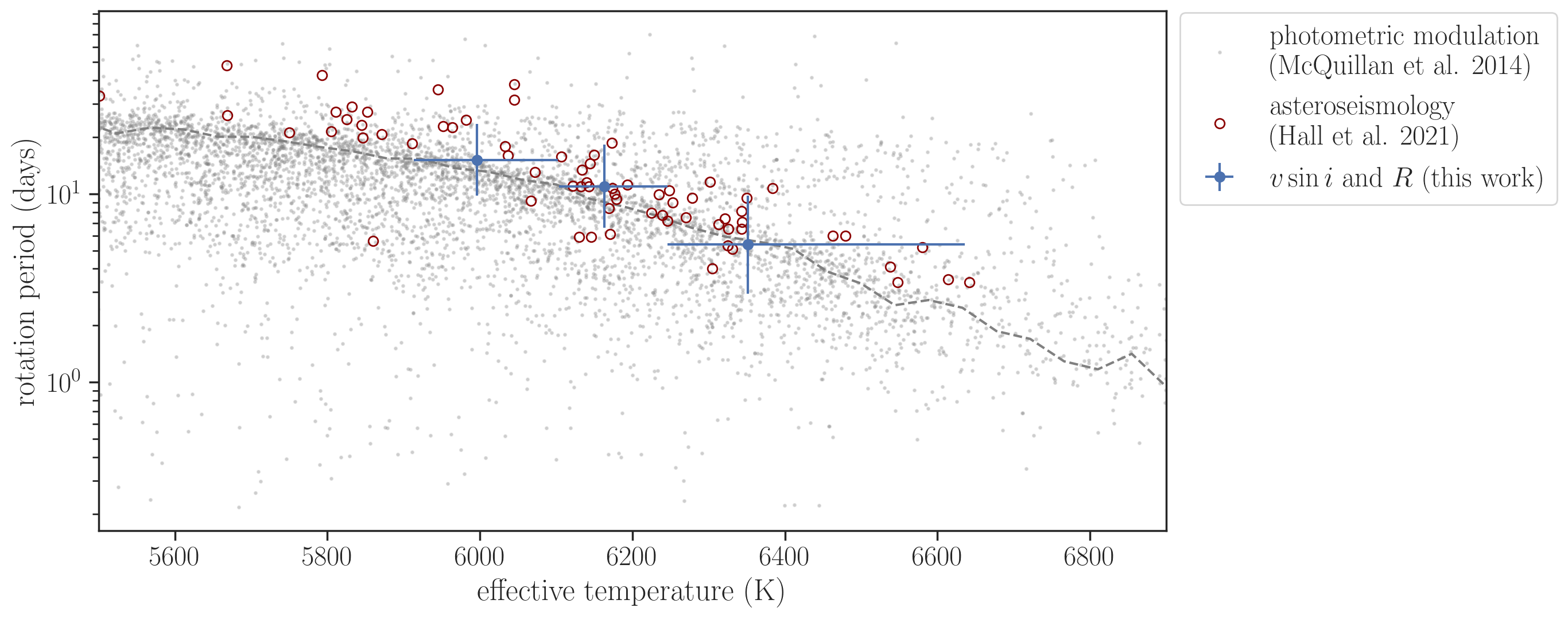}
    \caption{The rotation periods versus effective temperatures of the \textit{Kepler} stars. 
    Gray dots show the measurements for generic \textit{Kepler} stars using photometric modulation \citep{2014ApJS..211...24M};
    only the stars with LAMOST DR6 spectroscopic parameters are shown and those with $\log g<3.9$ are excluded. 
    The gray dashed line connects the mid-points of the 25\% highest density intervals computed for $100\,\mathrm{K}$ bins around each temperature and traces the mode of the $\prot$ distribution as a function of $T_\mathrm{eff}$.
    Red open circles are asteroseismic measurements by \citet{2021NatAs...5..707H}: only the measurements deemed robust (Flag 0) are shown and error bars are omitted for clarity.
    Blue filled circles with error bars are the mean and standard deviation (vertical error bars) estimated from our $v\sin i$ model for three different temperature bins (horizontal error bars; horizontal positions of the circles show the median $T_\mathrm{eff}$ in each bin); these are the values reported in Figure \ref{fig:upper3_temp}.}
    \label{fig:prot_teff}
\end{figure*}

In Figure \ref{fig:upper3_temp}, we split the entire sample into (i) stars with $T_\mathrm{eff}<6103.04\,\mathrm{K}$, (ii) stars with $6103.04\,\mathrm{K}<T_\mathrm{eff}<6245.96\,\mathrm{K}$, and (iii) stars with $T_\mathrm{eff}>6245.96\,\mathrm{K}$, and inferred the period distribution separately for each subsample. The threshold temperatures were chosen so that we have equal numbers of stars in each range. We find that hotter stars rotate more rapidly, as expected. We also find that the rotation periods of hotter stars show a larger dispersion. This could indicate that their rotation periods have not converged due to weaker magnetic braking, or could simply be due to a larger $T_{\rm eff}$ bin size.
We again find the overprediction of small $v\sin i$ stars in each subsample; this seems to argue against the hypothesis that the tension is due to our assumption that $p(x|\alpha)$ is common for all stars in the sample.
We also confirmed that the inferred distribution does not depend on whether those measurements were treated as upper limits or not.

As seen in Figure \ref{fig:all-pdfs}, the statistical uncertainty of $\prot$ is dominated by unknown $\sin i$: $\sin i\approx 1$ is a priori most likely, but we see a long tail toward smaller $\sin i$ (i.e., shorter $\prot$). On the other hand, systematic errors in $v\sin i$ or $R$ scale the whole inferred distribution of $\prot \propto R/v\sin i$ upwards or downwards. The source of potentially largest systematics would come from the macroturbulence broadening assumed in estimating $v\sin i$. We evaluate its impact by consulting several proposed macroturbulence relations in Appendix \ref{sec:vmacro} and find that the effect on $\prot$ is likely less than about $10\%$. Systematic errors in $R$ due to uncertainties in the stellar models and fundamental scales for stellar parameters are likely smaller \citep{2020arXiv201207957T}. These potential systematics of $\lesssim 10\%$ level do not affect the following discussion.

\subsection{Comparison with Periods from Photometric Modulation} \label{ssec:kepler_photometric}

We cross-matched our sample with the \citet{2014ApJS..211...24M} catalog of stellar rotation periods measured using quasi-periodic photometric modulation in the \textit{Kepler} light curves. We found that the photometric rotation periods are detected for 41 stars (28\%) in our $v\sin i$ sample.
In Figure \ref{fig:prot_comparison}, the $\prot$ distribution for the 41 stars from photometric modulation (thick gray) is compared with that from $v\sin i$ and $R$ for our sample stars (thin blue) in the form of the cumulative distribution function (CDF). 
We find evidence that the sample of photometrically measured periods includes more rapid rotators: we performed the one-sample Kolmogorov-Smirnov (KS) test and found the $p$-value of $0.01$ for the hypothesis that photometric periods are drawn from the mean distribution inferred from $v\sin i$.

\begin{figure}
	\includegraphics[width=\columnwidth]{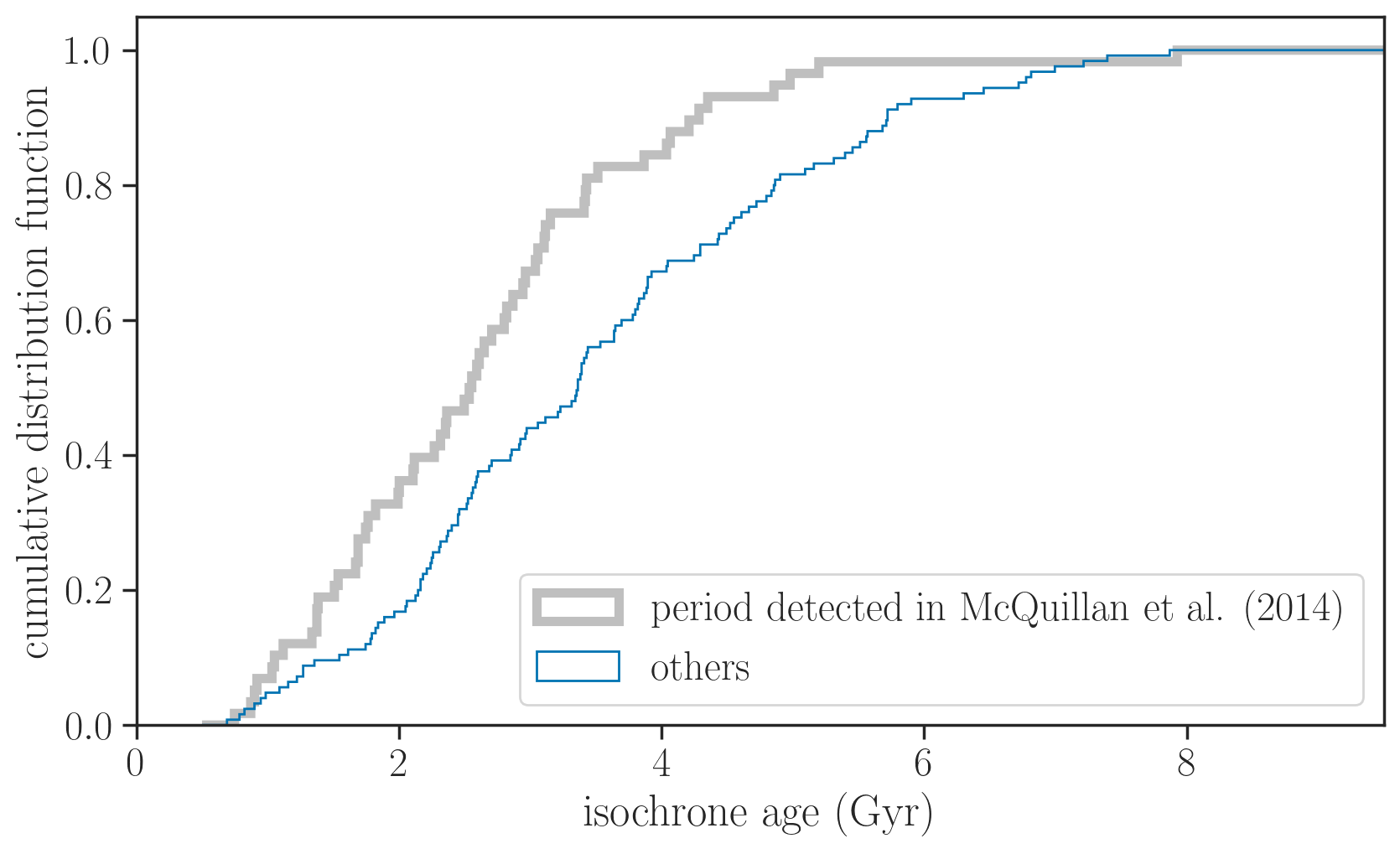}
    \caption{Isochrone-based ages of our sample stars with (thick gray) and without (thin blue) $\prot$ detections from photometric modulation \citep{2014ApJS..211...24M}.}
    \label{fig:age_cdf}
\end{figure}

The same difference is also seen in Figure \ref{fig:prot_teff}, where the photometric rotation periods for the entire sample of \citet{2014ApJS..211...24M} are plotted as a function of effective temperatures based on low-resolution spectroscopy from the Large Sky Area Multi-Object Fibre Spectroscopic Telescope (LAMOST) project \citep[DR6;][]{2012RAA....12.1197C,2014IAUS..306..340W,2015RAA....15.1095L,2016ApJS..225...28R}. The figure shows that photometric periods (gray dots) scatter down to shorter periods than typical values inferred from $v\sin i$ for the three temperature bins (blue filled circles with error bars; the same as in Figure \ref{fig:upper3_temp}).
The excess of rapid rotators qualitatively makes sense: they tend to exhibit larger photometric variabilities and so are easier to detect. Indeed, the isochrone ages of our sample stars indicate that the subset of them with $\prot$ detected in \citet{2014ApJS..211...24M} are significantly younger than the others, with the two-sample KS $p$-value of $10^{-3}$ (Figure \ref{fig:age_cdf}).

Despite the difference at small $\prot$, the distributions at longer $\prot$ are in fact similar. Figure \ref{fig:prot_teff} shows that the periods inferred from $v\sin i$ as a function of $T_\mathrm{eff}$ agree with the most typical $\prot$ inferred from photometric modulation in each bin and trace the ``long-period edge'' in the $\prot$--$T_\mathrm{eff}$ plane.\footnote{Our view of the ``long-period edge'' may be slightly different from the one in, e.g., \citet{2019ApJ...872..128V}. With the $T_\mathrm{eff}$ values from LAMOST, we find that $\prot$ distributions at different $T_\mathrm{eff}$ have better defined peaks at longer $\prot$ than seen when the  $T_\mathrm{eff}$ from the \textit{Kepler} input catalog is adopted. The dashed line in Figure \ref{fig:prot_teff} traces this ``ridge.''
} 
Namely, our inferred periods match the gray dashed line in Figure \ref{fig:prot_teff}, which connects the mid-points of the 25\% highest density intervals of $\prot$ computed for $100\,\mathrm{K}$ bins around each temperature and so is designed to trace the mode of the $\prot$ distribution as a function of $T_\mathrm{eff}$. We also see this in Figure \ref{fig:prot_comparison} where the gray and blue CDFs have the largest slopes (i.e., largest probability densities) at similar $\logp$ values of $\approx 1.1$.
In other words, typical $\prot$ of the stars {\it without} detected photometric modulation (i.e., $\approx 70\%$ of the sample) are similar to $\prot$ of those with detected modulation save for the most rapid rotators, and our results argue against the presence of a large number of stars rotating slower than those with detected photometric modulation.

\begin{figure*}
	\includegraphics[width=1.8\columnwidth]{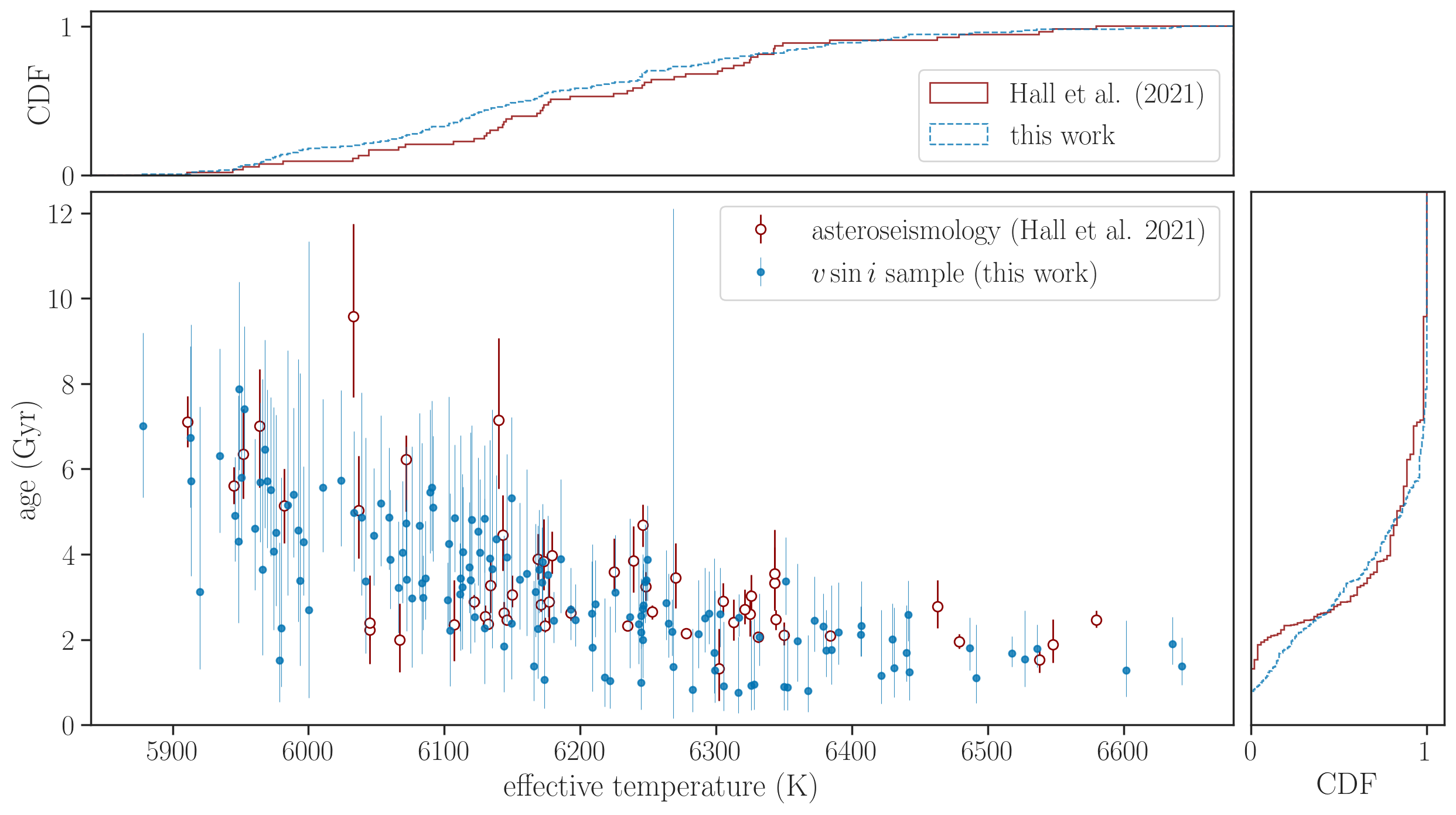}
    \caption{Ages versus effective temperatures of the asteroseismic sample \citep[red open circles and solid lines;][]{2021NatAs...5..707H} and of our $v\sin i$ sample (blue filled circles and dashed lines; from spectroscopy and isochrone fitting). The effective temperatures have $\sim100\,\mathrm{K}$ uncertainties including systematics.}
    \label{fig:age_teff}
\end{figure*}

\subsection{Comparison with Periods from Asteroseismology}\label{ssec:kepler_astero}

We also compare the results with asterosesmic measurements recently reported by \citet{2021NatAs...5..707H}. From their sample, we chose the stars with $5900\,\mathrm{K}<T_\mathrm{eff}<6600\,\mathrm{K}$ and $\log g>3.9$ that match our sample for comparison. 
Only the measurements that were deemed robust in their work (Flag 0) were considered.
These criteria left us with 53 stars from \citet{2021NatAs...5..707H}, in which seven stars are common with our $v\sin i$ sample.
Despite the different selection criteria, we find that the effective temperatures, masses, and ages of the asteroseismic stars chosen this way are similar to those of our sample based on spectroscopy and isochrone fitting (see Section \ref{ssec:kepler_sample}): we found the two-sample KS $p$-values of $\approx 0.1$ for these parameters (Figure \ref{fig:age_teff}). 

We find that the period distribution of the asteroseismic sample (thick red line in Figure \ref{fig:prot_comparison}) agrees better with our $v\sin i$-based result than that from photometric modulation. The hypothesis that the asteroseismic measurements are drawn from the $v\sin i$-based mean distribution is not rejected with the one-sample KS $p$-value of $0.2$. The distribution on the $\prot$--$T_\mathrm{eff}$ plane is also consistent with our estimates using $v\sin i$, perhaps except for the coolest bin (Figure \ref{fig:prot_teff}). 
The slightly shorter $\prot$ estimated from $v\sin i$ may reflect that the coolest stars in the asteroseismic sample are systematically older than ours (Figure \ref{fig:age_teff}). 
This could be because the asteroseismic oscillations are more likely detected for more evolved stars and/or older and magnetically less active stars \citep{2019FrASS...6...46M}.

\subsection{Evidence of Weakened Magnetic Braking}

\citet{2019ApJ...872..128V} found that the long-period edge of the $\prot$--$T_\mathrm{eff}$ distribution of the \textit{Kepler} stars from \citet{2014ApJS..211...24M} --- as also shown in our Figure \ref{fig:prot_teff} --- 
is not well reproduced by their forward model that accounts for a galactic stellar population, rotational evolution, and selection function of the \textit{Kepler} targets: their model predicted too many hot dwarfs rotating more slowly than observed with photometric modulation. They thus hypothesized that either the slowest rotators with the largest Rossby numbers have been missed in the photometric sample due to weakened activity and/or a transition in the morphology of surface active regions, or that the magnetic braking stalls around a critical Rossby number after which stars do not spin down so much as assumed in the standard model \citep[weakened magnetic braking;][]{2016Natur.529..181V}. 
The latter scenario has also been shown to explain the tension that some stars older than the middle of their main-sequence lifetimes rotate more rapidly than predicted from gyrochronology relations calibrated for younger stars \citep{2015MNRAS.450.1787A, 2016Natur.529..181V}, although the critical Rossby number at which this happens or the necessity of the modified braking law itself are still under debate \citep[e.g.][]{2019MNRAS.485L..68L,2019ApJ...871...39M}.
The agreement we found in Section \ref{ssec:kepler_photometric} between the typical $\prot$ from photometric modulation and that from $v\sin i$ --- which is free from the bias associated with detectability of the photoemtric modulation --- 
argues against the presence of many slow rotators that were missed in the photometric modulation sample and supports the weakened magnetic braking interpretation.

The above argument is rather qualitative.
As an attempt to test the stalled spin-down scenario more directly, we used our isochrone parameters to pick up the 36 stars in our sample with $1.05\,M_\odot$--$1.15\,M_\odot$, corresponding to Figure 3 of \citet{2016Natur.529..181V} with the ZAMS $T_\mathrm{eff}$ of $5900\,\mathrm{K}$--$6200\,\mathrm{K}$. We divided them into the younger and older subsamples with roughly uniform age distributions between $1$--$4.5\,\mathrm{Gyr}$ and $4.5$--$7\,\mathrm{Gyr}$, and derived $\prot$ distribution for each of them.
We found the mean $\prot$ of $11.7\,\mathrm{days}$ and $14.5\,\mathrm{days}$ for the younger and older subsamples, respectively, which roughly match the evolution tracks with weakend magnetic braking in Figure 3 of \citet{2016Natur.529..181V}. That said, the inferred standard deviations of $\logp$ are $\approx 0.28$, and the results are based on only 18 stars in each subsample; a larger number of stars is required to obtain a more decisive conclusion.

\begin{figure*}
	\includegraphics[width=1.8\columnwidth]{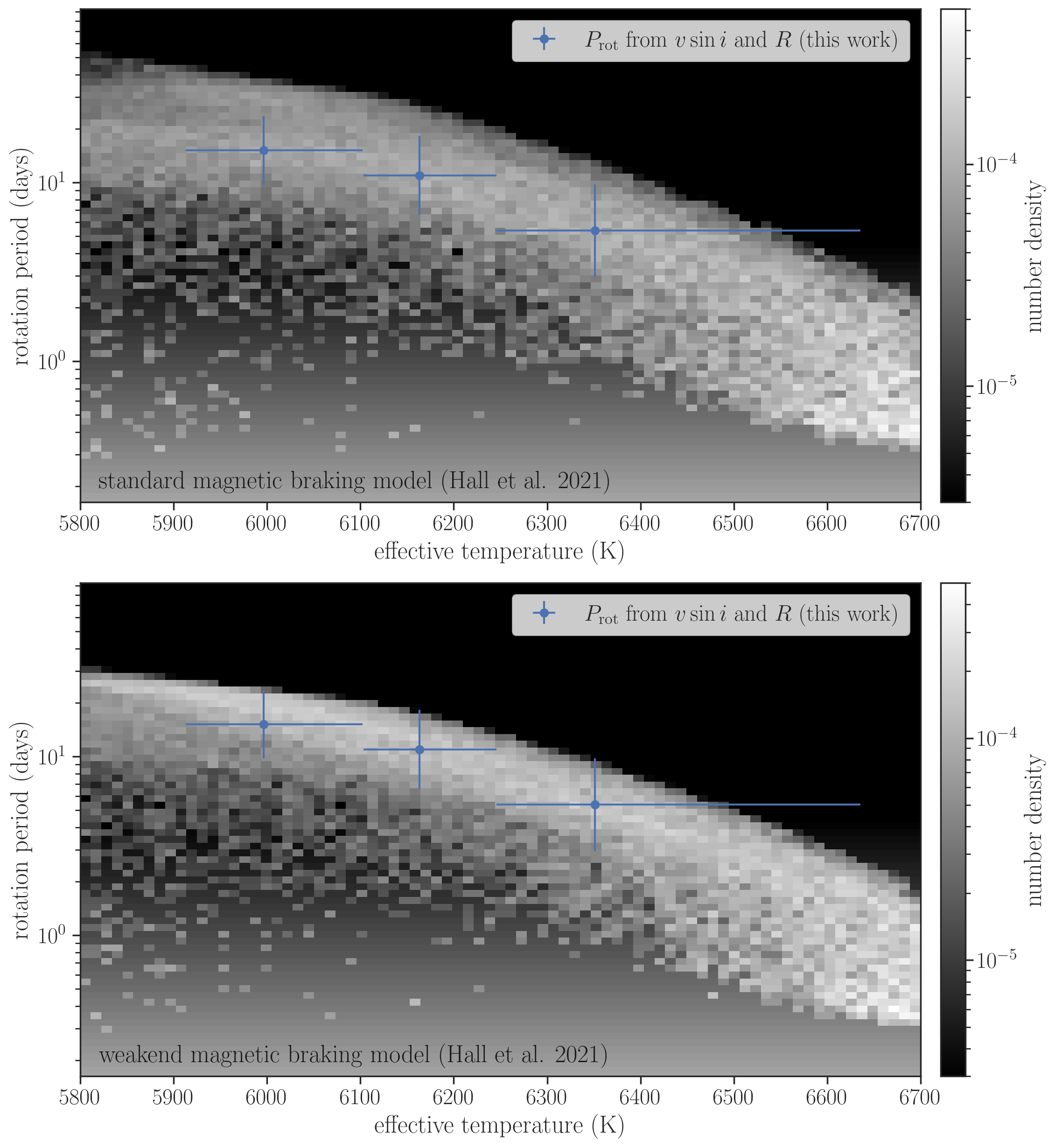}
    \caption{The rotation periods versus effective temperatures estimated from our $v\sin i$ model (blue filled circles with error bars) compared with the stellar population models evolved under standard (top) and weakened (bottom) magnetic braking models from \citet{2021NatAs...5..707H}. 
    The vertical positions of the circles and error bars show the mean and standard deviation of the $\prot$ distributions, respectively. The horizontal positions and error bars show the medians and widths of the $T_\mathrm{eff}$ bins (cf. Figure \ref{fig:upper3_temp}).
    The model populations are shown as a density plot, where the model stars with $\log g<3.9$ are excluded so that they match our sample.}
    \label{fig:prot_teff_model}
\end{figure*}

More recently, \citet{2021NatAs...5..707H} leveraged their precise asteroseismic constraints on the parameters of individual stars (in particular the rotation periods and ages) to perform a model comparison
between the standard and weakened magnetic braking laws, and concluded that the latter is significantly favored.
This conclusion is mainly driven by the lack of slowly rotating main-sequence stars with $T_\mathrm{eff} < 6250\,\mathrm{K}$ that are expected to be present in the standard braking model, as shown in Figure 3 of \citet{2021NatAs...5..707H}. In Figure \ref{fig:prot_teff_model}, we compare our estimates (mean and standard deviation) of $\prot$ against the stellar population models from \citet{2021NatAs...5..707H} and find that the same is true. 
This agreement provides further evidence for the breakdown of the standard magnetic braking law for stars beyond the middle of their main-sequence lifetimes.

We attempt to quantify this comparison by evaluating the Bayesian evidence for the entire data for all the sample stars $\mathcal{D}$:\footnote{This $\mathcal{D}$ does not only include the data for $R$ and $v\sin i$ but also those used for isochrone fitting.}
\begin{align}
    \notag
    p(\mathcal{D}|\mathcal{M}) &= \prod_j p(\mathcal{D}_j|\mathcal{M})
    = \prod_j \int p(\mathcal{D}_j|\theta)\,p(\theta|\mathcal{M})\,\mathrm{d}\theta\\
    &\approx \prod_j \left[ {1 \over K}\sum_k
    p(\mathcal{D}_j|\theta_{k}) \right],
    \quad \theta_{k} \sim p(\theta|\mathcal{M})
    \label{eq:evidence}
\end{align}
where
$\theta = (M_{\star}, \mathrm{[Fe/H]},  \mathrm{age}, P_{\mathrm{rot}})$.
Here the model $\mathcal{M}$ assumes either the standard ($\mathcal{S}$) or weakened ($\mathcal{W}$) magnetic braking law, and the samples $\theta_k$ are drawn from the model populations simulated by \citet{2021NatAs...5..707H} assuming these models.
The likelihood $p(\mathcal{D}_j|\theta)$ for the $j$th star was computed by assuming independent Gaussian likelihoods for $(M_{\star}, \mathrm{[Fe/H]}, \log(\mathrm{age}))$ and using Equation \ref{eq:margL} for $P_\mathrm{rot}$. Equation \ref{eq:evidence} yields the odds ratio of $p(\mathcal{D}|\mathcal{W})/p(\mathcal{D}|\mathcal{S})=\exp(3.9)=49$,\footnote{Note that this entails essentially the same information as used in \citet{2021NatAs...5..707H}. They evaluated $p(\mathcal{D}|Q_\mathrm{WMB})=\prod_j p(\mathcal{D}_j|Q_\mathrm{WMB})=\prod_j [Q_\mathrm{WMB}\,p(\mathcal{D}_j|\mathcal{W}) + (1-Q_\mathrm{WMB})\,p(\mathcal{D}_j|\mathcal{S})]$ and reported the posterior $p(Q_\mathrm{WMB}|\mathcal{D})$ adopting the prior uniform in $(0,1)$ for $Q_\mathrm{WMB}$.}
which implies that the weakened law is strongly favored over the standard law. 
We note that this comparison does not depend on the hyperprior $p(\prot|\alpha)$ used to infer the $\prot$ distribution, because the evaluation of $p(\mathcal{D}_j|\mathcal{M})$ involves only the likelihood for $P_{\mathrm{rot},j}$.

\begin{figure*}
	\includegraphics[width=1.8\columnwidth]{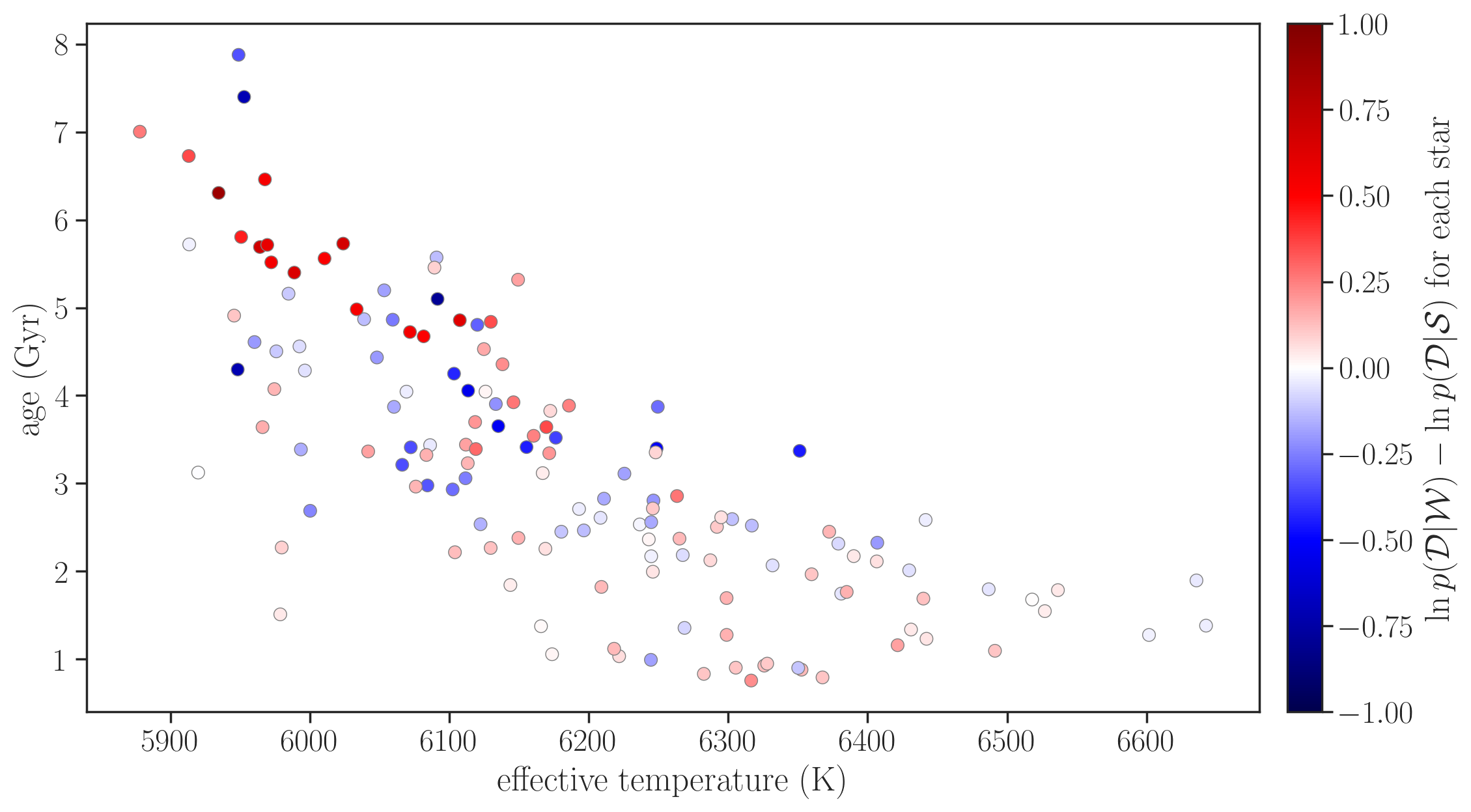}
    \caption{Isochrone ages and spectroscopic effective temperatures of our sample stars (same as in Figure \ref{fig:age_teff}), color coded with the logarithm of the odds ratio $\ln \left[p(\mathcal{D}_j|\mathcal{W})/p(\mathcal{D}_j|\mathcal{S})\right]$ between the weakened and standard braking models from \citet{2021NatAs...5..707H}.}
    \label{fig:age_teff_evidence}
\end{figure*}

The values of $\ln \left[p(\mathcal{D}_j|\mathcal{W})/p(\mathcal{D}_j|\mathcal{S})\right]$ for each star in Figure \ref{fig:age_teff_evidence} show that the evidence in favor of the weakened braking model $\mathcal{W}$ comes from the stars in the top-left part of the diagram, i.e., the coolest and oldest stars in the sample. On the one hand, this is the signature expected from the weakened braking law, in which only stars beyond the middle of their main-sequence lifetimes exhibit significant deviation from the standard braking law. On the other hand, this also implies that the ``population-level'' result is driven by a relatively small number of stars in the sample.
Another generic note is that this is a relative comparison between the two models, which itself cannot evaluate whether the model with the weakened magnetic braking is a suitable one for the data. Indeed, we find that both population models include more rapid rotators than we inferred espeically at higher $T_\mathrm{eff}$, as also seen in Figure 3 of \citet{2021NatAs...5..707H}; so even the model with weakened braking does not explain all the properties of the sample. The same caveats also apply to the analysis of \citet{2021NatAs...5..707H}.

Here we emphasize again that our results fully take into account the stars with $v\sin i$ below the detection limit. The sample is constructed without excluding such stars, and our method treats them consistently with measurements well above the detection limit. Although we did find evidence that low values of $v\sin i \lesssim 3\,\mathrm{km/s}$ are affected by systematic errors in Section \ref{ssec:kepler_results}, there we also showed that the recovered $\prot$ distribution was insensitive to those measurements mainly because the fraction of such stars is small ($\approx 10\%$). 
We also note that not all of them are real slow rotators. For example, if we crudely assume that all sample stars have $v=2\pi R/\prot \sim 9\,\mathrm{km/s}$ (mean of our inferred value), $\approx 6\%$ should have $\sin i < 1/3$ or $v\sin i <3\,\mathrm{km/s}$ for isotropically oriented spins (i.e., uniformly distributed $\cos i$). Therefore there is not much room to ``hide'' slow rotators among those stars with $v\sin i$ below the detection limit. 

\section{Summary and Conclusion} \label{sec:summary}

We presented a probabilistic framework to infer the rotation period distribution for a sample of stars with isotropically oriented spin axes, based on the data sets that constrain $v\sin i$ and radii. This $v\sin i$-based inference is complementary to the period measurements based on quasi-periodic brightness modulation in the photometric light curves, because the latter method is only sensitive to stars that exhibit large photometric variabilities and may be biased against less active stars. Our method was tested using simulated data sets and was shown to work well for $\gtrsim 100$ stars with current best achievable constraints, unless the true rotation period distribution includes sharp discontinuities. 
For data sets with different sizes and qualities, similar tests may be mandatory to adjust the priors on the hyperparameters to obtain sensible results.
We also note that our choice for the hyperprior is just one example that turned out to work reasonably well for the present problem. For example, one could use frameworks other than a Gaussian process (e.g., neural network) to construct the hyperprior $p(x|\alpha)$. Alternatively, one could adopt simpler, physically motivated parametric distributions for $p(x|\alpha)$ if exist, which greatly reduce the number of model parameters and simplify the sampling process. 

We applied our method to the late-F/early-G \textit{Kepler} stars observed with Keck/HIRES to infer their rotation period distribution, in a manner that is not biased against stars with weak photometric variabilities.
We found that the typical rotation periods of these stars are similar to the values inferred from photometric modulation, although we found evidence that younger, rapidly rotating stars are overrepresented in the photometric sample.
We also found a good agreement between our inference and the periods measured for similar \textit{Kepler} stars with asteroseismology, including their dependence on the effective temperatures.
These comparisons demonstrate the usefulness of the $v\sin i$-based method as an alternative diagnosis of stellar rotation, and support the 
weakened magnetic braking hypothesis that the standard braking law breaks down at least for some hot stars older than the middle of their main-sequence lifetimes.

\section*{Acknowledgements}

We thank BJ Fulton, Howard Isaacson, and Andrew Howard for curating and distributing the HIRES spectra studied here. We also appreciate the referee's constructive comments which helped to improve the manuscript.
This work has made use of data from the European Space Agency (ESA) mission
{\it Gaia} (\url{https://www.cosmos.esa.int/gaia}), processed by the {\it Gaia}
Data Processing and Analysis Consortium (DPAC, \url{https://www.cosmos.esa.int/web/gaia/dpac/consortium}). Funding for the DPAC
has been provided by national institutions, in particular the institutions
participating in the {\it Gaia} Multilateral Agreement.
K.M. acknowledges the support by JSPS KAKENHI Grant Number 21K13980.
E.A.P. acknowledges the support of the Alfred P. Sloan Foundation.

\section*{Data Availability}

The data and the code underlying this article are available from \url{https://github.com/kemasuda/prot_from_vsini}.



\bibliographystyle{mnras}
\bibliography{main} 




\appendix

\section{Formal Derivation of the Hierarchical Model} \label{sec:derivation}

Here we derive the equations presented in Section \ref{ssec:method_framework} more formally, starting from the full Bayesian solution of the problem. See, e.g., \cite{2010ApJ...725.2166H} and \citet{2014ApJ...795...64F} for different applications and pedagogical introductions of the method.

We have $N$ stars labeled by $j$. For each star we have the data $D_j$ 
that depend on three parameters: log-period $x_j$,  radius $R_j$, and cosine of inclination $\cos i_j$. By $\bm{y}$ we denote a set of $y_j$ for all stars: $\bm{D}=\{D_j\}_{j=1}^N$, $\bm{x}=\{x_j\}_{j=1}^N$, and so on. We assume that the data sets in different systems are independent and that the likelihood satisfies
\begin{align}
    p(\bm{D}|\bm{x}, \bm{R}, \bm{\cos i})
\label{eq:assumption_likelihood}
    = \prod_{j=1}^N p(D_j|x_j,  R_j, \cos i_j).
\end{align}

We assume separable priors for all $R_j$ and $\cos i_j$. For $x_j$, we assume that each $x_j$ follows the common hyperprior $p(x|\alpha)$, where $\alpha$ denotes the set of hyperparameters ($\alpha$, $a$, and $s$ in the text): so $x$'s of different stars are independent only when conditioned on $\alpha$. Then the full prior PDF is:
\begin{align}
\notag
    &p(\bm{x}, \bm{R}, \bm{\cos i}, \alpha)\\
\notag
    &= p(\bm{R})\,p(\bm{\cos i})\,p(\bm{x}|\alpha)\,p(\alpha)\\
\label{eq:assumption_prior}
    &= \left[\prod_{j=1}^N p(R_j)\,p(\cos i_j)\,p(x_j|\alpha)\right]\,p(\alpha).
\end{align}
The hyperprior $p(x|\alpha)$ represents the underlying distribution of $x$ in the sample which we wish to infer. 

The above assumptions yield the full joint posterior PDF conditioned on all data as:
\begin{align}
\notag
    &p(\bm{x}, \bm{R}, \bm{\cos i}, \alpha|\bm{D})\\
\notag
    &={1\over p(\bm{D})}\, p(\bm{D}|\bm{x}, \bm{R}, \bm{\cos i}, \alpha)\,p(\bm{x}, \bm{R}, \bm{\cos i}, \alpha)\\
\notag
    &={1\over p(\bm{D})}\left[\prod_{j=1}^N p(D_j|x_j, R_j, \cos i_j)\,p(R_j)\,p(\cos i_j)\,p(x_j|\alpha)\right]
    \,p(\alpha)\\
    &={1\over p(\bm{D})}\left[\prod_{j=1}^N p(D_j, R_j, \cos i_j|x_j)\,p(x_j|\alpha)\right]
    \,p(\alpha),
\end{align}
where we used Bayes' theorem in the first equality, Equations~\ref{eq:assumption_likelihood} and \ref{eq:assumption_prior} in the second, and the definition of conditional probability in the third.
In principle, it is possible to sample directly from this posterior PDF, but that involves many parameters to sample when $N$ is large.
In this paper, we did not deal with individual $\bm{x}$, $\bm{\cos i}$, and $\bm{R}$ but marginalized them out to obtain
\begin{align}
\notag
    &p(\alpha|\bm{D}) = \int p(\bm{x}, \bm{R}, \bm{\cos i}, \alpha|\bm{D})\,\mathrm{d}\bm{x}\,\mathrm{d}\bm{R}\,\mathrm{d}\bm{\cos i}\\
\notag
    &= {1\over p(\bm{D})}\,\prod_{j=1}^N \left[\int p(D_j, R_j, \cos i_j | x_j)\,\mathrm{d}R_j\,\mathrm{d}\cos i_j \,p(x_j|\alpha)\,\mathrm{d}x_j \right]\,p(\alpha)\\
    &= {1\over p(\bm{D})}\,\left[ \prod_{j=1}^N \int p(D_j|x_j)\,p(x_j|\alpha)\,\mathrm{d}x_j \right]\,p(\alpha).
\end{align}
The product in the last line within the square brackets is $p(\bm{D}|\alpha)$ in Equation~\ref{eq:likelihood}.

Let us then consider $x$ of the $(N+1)$th star, $X$, for which we do not yet have data --- what will be the PDF of $X$ given all the existing data $\bm{D}$? That is the distribution of $x$ we inferred from the population, and 
is given by the following marginalization similar to the one above:
\begin{align}
    \notag
    &p(X|\bm{D}) = \int p(X, \bm{x}, \bm{R}, \bm{\cos i}, \alpha|\bm{D})\,\,\mathrm{d}\bm{x}\,\mathrm{d}\bm{R}\,\mathrm{d}\bm{\cos i}\,\mathrm{d}\alpha\\
    \notag
    &= {1\over p(\bm{D})} \int \left[\prod_{j=1}^N \int p(D_j|x_j)\,p(x_j|\alpha)\,\mathrm{d}x_j\right]\,p(X|\alpha)\,p(\alpha)\,\mathrm{d}\alpha\\
    \notag
    &= \int p(X|\alpha)\,{p(\bm{D}|\alpha)\,p(\alpha)\over p(\bm{D})}\,\mathrm{d}\alpha\\
    &= \int p(X|\alpha)\,p(\alpha|\bm{D})\,\mathrm{d}\alpha.
\end{align}
This is Equation~\ref{eq:meanposterior}.

In this paper, we further assume that $D_j$ is separated into two independent subsets $D_{R,j}$ and $D_{u,j}$ that depend only on $R$ and $u \equiv v\sin i$, respectively.\footnote{
This is a simplifying assumption: $R$ and $v\sin i$ both depend on the same high-resolution spectrum (because the former depends on $T_\mathrm{eff}$) and so could be dependent.
}  Then we have
\begin{align}
\notag
    &p(D_j|x_j, R_j, \cos i_j)\\
\notag
    &=p(D_{R,j}|x_j, R_j, \cos i_j)\,p(D_{u,j}|x_j, R_j, \cos i_j)\\
    &= p(D_{R,j}|R_j)\,p(D_{u,j}|x_j, R_j, \cos i_j),
\end{align}
and the marginal likelihood of $x_j$, $p(D_j|x_j)$, is computed as
\begin{align}
\notag
    &p(D_j|x_j) = \int p(D_j, R_j, \cos i_j|x_j) \,\mathrm{d}R_j\,\mathrm{d}\cos i_j\\
\notag
    &= \int p(D_j|x_j, R_j, \cos i_j)\,p(R_j)\,p(\cos i_j)\,\mathrm{d}R_j\,\mathrm{d}\cos i_j\\
    &= \int p(D_{R,j}|R_j)\,p(D_{u,j}|x_j, R_j, \cos i_j)\,p(R_j)\,p(\cos i_j)\, \mathrm{d}R_j\,\mathrm{d}\cos i_j.
\end{align}
This is Equation~\ref{eq:margL}.

\section{Note on the Stellar Parameters}\label{sec:logg}

\begin{figure}
	\includegraphics[width=1.\columnwidth]{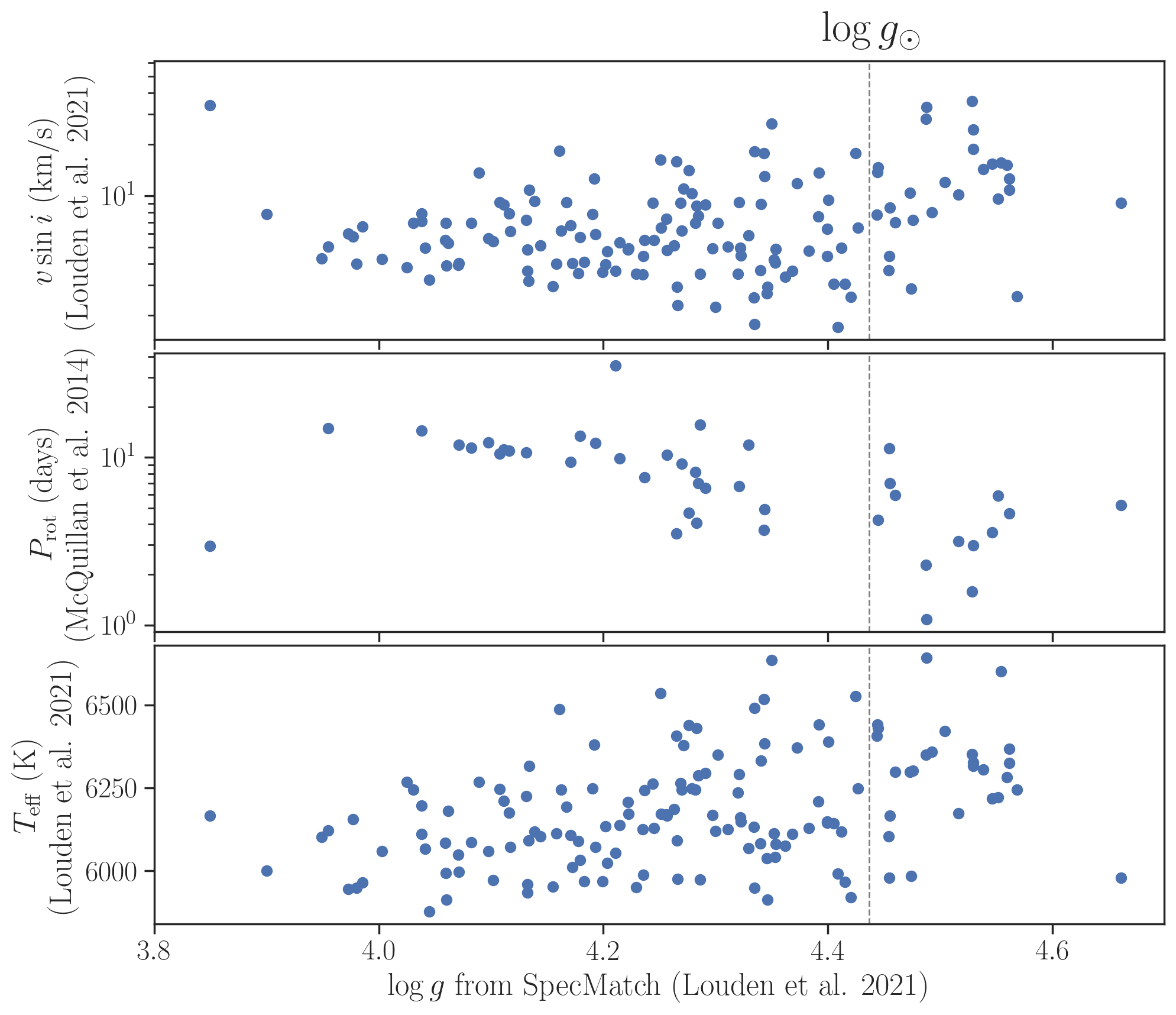}
    \caption{Some stars show unreasonably high $\log g$, which is also associated with rapid rotation and high effective temperature.}
    \label{fig:smsyn_logg}
\end{figure}

Although the stars in \citet{2021AJ....161...68L} are all hotter than the Sun, {\tt SpecMatch-Syn} assigned $\log g$ larger than the solar value for some of them. 
Their parallaxes and absolute magnitudes do not support that these stars are more compact than the Sun, and so these $\log g$ values are most likely overestimated.
We find that those stars tend to be hot stars with rapid rotation (Figure \ref{fig:smsyn_logg}) based on {\tt SpecMatch-Syn} parameters and rotation periods measured with photometric modulation by \citet{2014ApJS..211...24M}.
This might suggest that $v\sin i$, which is also associated with broadening of absorption lines and could be correlated with $\log g$, is biased.

This finding motivated us to adopt external information on $\log g$ in the spectrum fitting. 
We estimated $\log g$ independent from spectroscopy by fitting MIST stellar models to the 2MASS $JHK_\mathrm{s}$ photometry adopting the \textit{Gaia} EDR3 distance \citep{2021AJ....161..147B} of each star. The $\log g$ values thus derived were found to show an excellent agreement with those from asteroseismology for the stars in common with the sample of \cite{2021NatAs...5..707H}: the mean difference is 0.00 and standard deviation is 0.03.
We then re-ran the {\tt SpecMatch-Syn} algorithm on our sample stars adopting those constraints on $\log g$ assuming a uniform uncertainty of $0.05$ and found that the mean change in $v\sin i$ is $0.07\,\mathrm{km/s}$ with the standard deviation of $0.07\,\mathrm{km/s}$. Therefore, this potential systematic error in $\log g$ does not bias $v\sin i$.

\section{Dependence on the Macroturbulence Relation} \label{sec:vmacro}

\begin{figure*}
	\includegraphics[width=1.26\columnwidth]{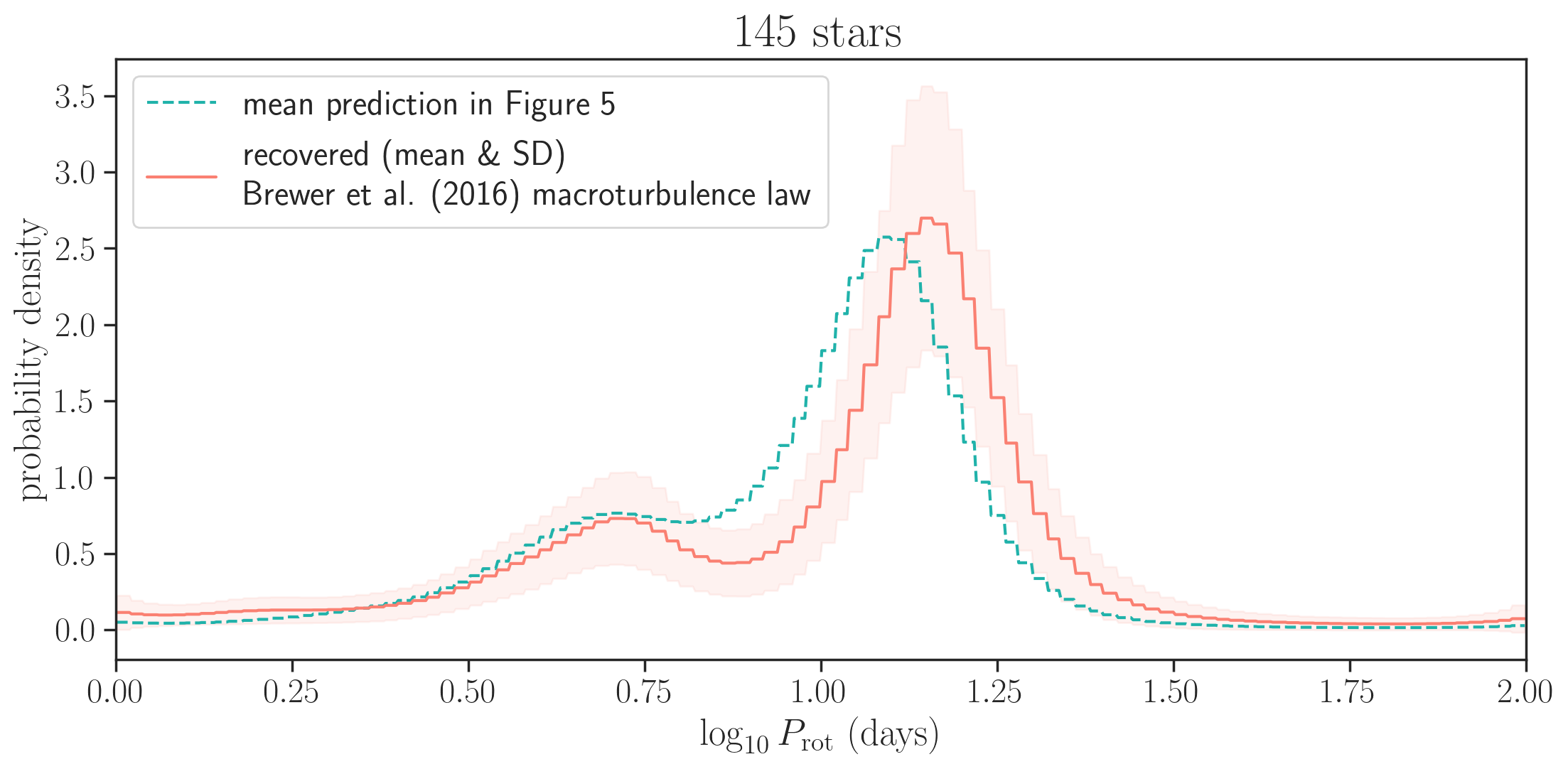}
	\includegraphics[width=0.81\columnwidth]{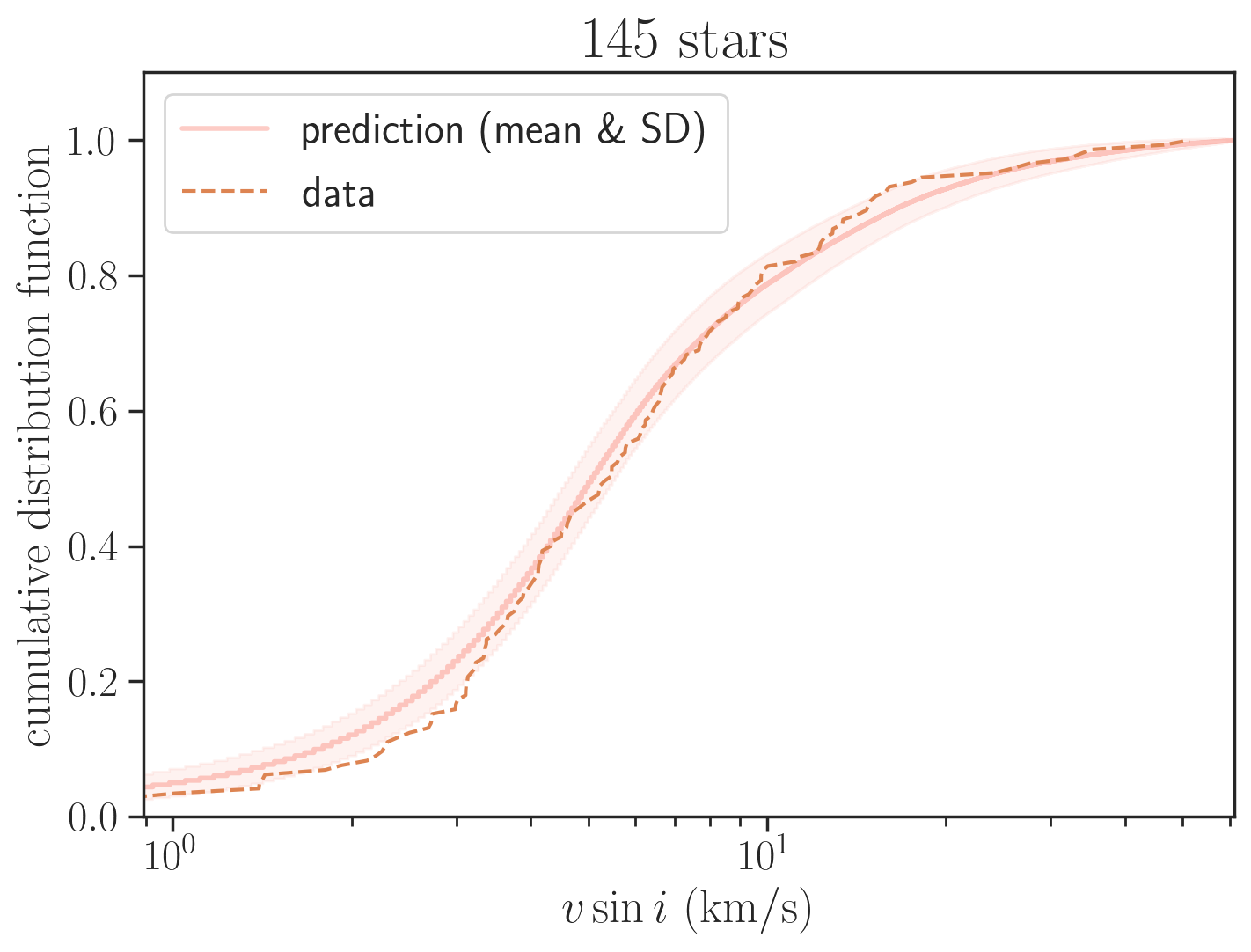}
    \caption{(Left) The $\log_{10} P$ distribution inferred for the whole sample, adopting the macroturbulence law derived by \citet{2016ApJS..225...32B} for dwarfs. The green dashed line shows the mean prediction in Figure \ref{fig:all}. (Right) Predicted $v\sin i$ distribution compared against the data.}
    \label{fig:all_b16macro}
\end{figure*}

We attempt to evaluate potential systematic errors associated with line broadening due to macroturbulence.
{\tt SpecMatch-Syn} estimates $v \sin i$ adopting the isotropic radial--tangential model for macroturbulence \citep{2005oasp.book.....G}.
The macroturbulence velocity dispersion is computed as a function of $T_\mathrm{eff}$ using the empirical relation derived by \citet{2005ApJ...622.1102F}. The relation was obtained by first
fitting line profiles fixing $v\sin i=0$ to derive upper limits on the macroturbulence dispersion, and then by fitting their lower envelope as a function of $T_\mathrm{eff}$.
Other empirical relations between the macroturbuelence velocity dispersion and $T_\mathrm{eff}$ have been proposed for hot stars relevant to our sample \citep[see also][]{1997MNRAS.284..803S}: 
\citet{1984ApJ...281..719G} modeled line profiles of very high-resolution spectra to disentangle $v\sin i$ and macroturbulence; \citet{2014MNRAS.444.3592D} fitted macroturbulence fixing $v\sin i$ derived from asteroseismology; \citet{2016ApJS..225...32B} updated the \citet{2005ApJ...622.1102F} relation using the same method.
For the temperature range of our interest, the \citet{2005ApJ...622.1102F} relation adopted in {\tt SpecMatch-Syn} predicts among the smallest macroturbulence dispersions, while the \citet{2016ApJS..225...32B} relation for dwarfs predicts the largest values, by up to a few $\mathrm{km/s}$ compared to the \citet{2005ApJ...622.1102F} relation.
Although \citet{2016ApJS..225...32B} do not appear to intend that the relation be used for such hot stars given the small number of calibration stars (cf. their Figure 10), we re-ran the whole analysis in Section \ref{sec:kepler} adopting the relation in \citet{2016ApJS..225...32B} for dwarfs to gauge the maximum possible impact due to systematics in the macroturbulence law.

The results are shown in Figure \ref{fig:all_b16macro}. The right peak in the distribution shifts by $\approx 10\%$ toward longer values, since $v\sin i$ becomes smaller for the larger adopted macroturbulence velocities. 
We find the mean and standard deviation of $\logp$ to be $1.01\pm0.03$ and $0.31\pm0.03$, respectively, compared to $0.97$ and $0.27$ from our fiducial analysis (Section \ref{sec:kepler}).
As a function of $T_\mathrm{eff}$, the result from the highest temperature bin (i.e. largest $v\sin i$) remained largely unaffected, while $\prot$ in the cooler bins increased by $\approx 20\%$. The shifts are smaller than the inferred standard deviation in each bin.

Interestingly, the tension at smallest $v\sin i$ in our fiducial analysis disappeared (Figure \ref{fig:all_b16macro}, right).
However, the agreement with the asteroseismic sample became poorer at larger $v\sin i$: we found $-1.2\pm1.2\,\mathrm{km/s}$ (spectroscopic minus seismic) compared to $-0.3\pm0.9\,\mathrm{km/s}$ in our fiducial analysis. Thus the $v\sin i$ values of those stars instead appear to be underestimated due to too large macroturbulence and the tension remains unresolved. Given that the stars with $v\sin i>3\,\mathrm{km/s}$ are the majority of the sample and play a more important role in determining the period distribution, we do not consider this analysis based on the \citet{2016ApJS..225...32B} macroturbulence law to be more accurate and reliable than the one in Section \ref{sec:kepler}.


\bsp	
\label{lastpage}
\end{document}